\documentclass[epj]{svjour} 

\usepackage{graphicx}

\usepackage{color}

\usepackage{float}
\usepackage{pslatex}

\begin{document}

%
%
%

\title{Performance Studies of Prototype II for the
CASTOR forward Calorimeter at the CMS Experiment}

\author{ 
X.~Aslanoglou$^1$, 
N.~Bakirci$^2$,
S.~Cerci$^2$, 
A.~Cyz$^3$,
D.~d'Enterria$^4$
E.~Gladysz-Dziadus$^3$, 
L.~Gouskos$^5$, 
A.~Ivashkin$^6$,
C.~Kalfas$^7$,
P.~Katsas$^5$, 
A.~Kuznetsov$^8$, 
Y.~Musienko$^8$,
A.D.~Panagiotou$^5$,
E.~Vlassov$^9$,
}

\institute{
$^1$University of Ioannina, Hellas\\
$^2$University of Cukurova, Turkey\\
$^3$Institute of Nuclear Physics, Krakow, Poland\\
$^4$CERN, Geneva, Switzerland\\
$^5$University of Athens, Hellas\\
$^6$Institute of Nuclear Research, Moscow, Russia\\
$^7$NRC ``Demokritos'' INP, Hellas\\
$^8$Northeastern University, Boston, USA\\
$^9$ITEP, Moscow, Russia
}

\date{Received: date / Revised version: date}

\abstract{
We present results of the performance of the second prototype of the CASTOR 
quartz-tungsten sampling calorimeter, to be installed in the very forward region 
of the CMS experiment at the LHC. The energy linearity and resolution, as well 
as the spatial resolution of the prototype to electromagnetic and hadronic showers 
are studied with $E=$ 20-200 GeV electrons,  $E=$ 20-350 GeV pions, and $E=$ 50, 150 GeV 
muons from beam tests carried out at CERN/SPS in 2004. The responses of the 
calorimeter using two different types of photodetectors (avalanche photodiodes 
APDs, and photomultiplier tubes PMTs) are compared.
}

\titlerunning{Performance of Prototype II for the CMS CASTOR forward calorimeter}
\authorrunning{X.~Aslanoglou {\it et al.}}
\maketitle


\section{Introduction}

\begin{figure*}[htbp] 
\begin{center}
\includegraphics[width=13cm]{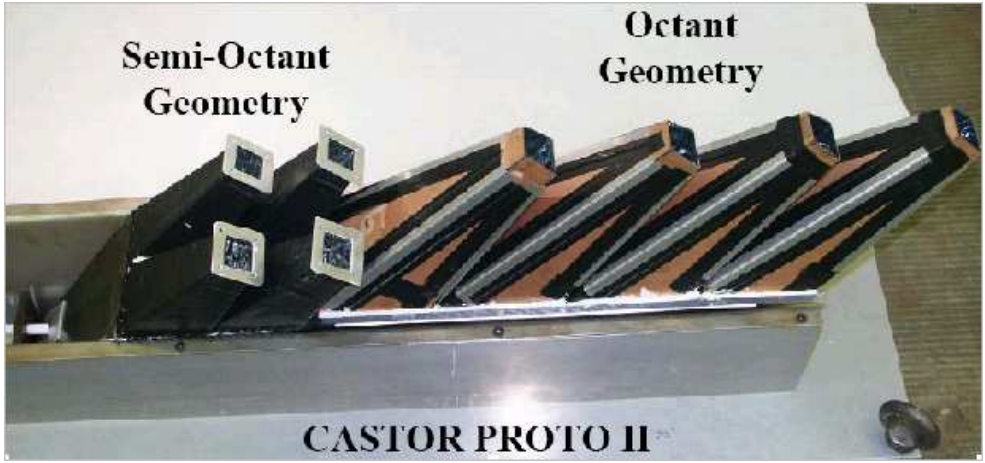}
\caption{Picture of the CASTOR prototype II calorimeter before assembling the photodetectors. 
The semi-octant geometry of the EM section (length: 14 cm) and the octant geometry of the 
HAD section (length: 40 cm) can be seen.}
\label{fig:proto2_pic}
\end{center}
\end{figure*}

The CASTOR (Centauro And Strange Object Research) detector is a quartz-tungsten 
sampling calorimeter, which has been proposed for the study of the very forward rapidity 
region in heavy ion and proton-proton collisions in the multi-TeV range at the 
LHC~\cite{castor}. Its main physics motivation is to complement the nucleus-nucleus physics 
programme, focused mainly in the baryon-free region at midrapidity~\cite{ptdr}. CASTOR will be 
installed in the CMS experiment at 14.38 m from the interaction point, covering the 
pseudorapidity range 5.2 $< \eta <$ 6.6 and will, thus, contribute not only to the 
heavy ion programme, but also to diffractive and low-$x$ physics in pp collisions~\cite{castor_forw}.
The results of the beam test and simulation studies with CASTOR prototype I~\cite{castor_protoI} 
prompted us to construct a second prototype using quartz plates, avalanche photodiodes (APDs) 
as well as photomultiplier tubes (PMTs), and air-core light-guides with inner reflective 
foil (Dupont polyester film reflector coated with AlO and reflection enhancing 
dielectric layer stack SiO$_2$+TiO$_2$). In addition, we tested a new semi-octant 
($\phi$ = 22.5$^\circ$) geometry of the readout unit in the electromagnetic section. 
The beam tests were carried out in the H2 line at the CERN SPS in 2004 using beams of 
electrons, pions and muons. The prototype II calorimeter consists of an electromagnetic 
(EM) and a hadronic (HAD) section, built in an octant sector (Fig.~\ref{fig:proto2_pic}). 
Both calorimeters are constructed with successive layers of tungsten plates as absorber 
and fused silica quartz plates as active medium. The EM part (14 cm length) is further 
divided into two semi-octant sectors and is longitudinally segmented into 2 sections, so 
that there are 4 independent readout units in total. The HAD part (40 cm length) 
retains the octant geometry of prototype I and is longitudinally segmented into 4 sections. 
The \v{C}erenkov light produced by the passage of relativistic particles through the 
quartz medium is collected in sections along the length of the calorimeters and focused 
by air-core light guides onto the photodetector devices, APDs or PMTs.

\section{Technical description}
\label{sec:WQ}

The CASTOR detector is a \v{C}erenkov-effect based calorimeter with tungsten absorber 
and quartz plates as active material. The main advantages of quartz calorimeters are their
radiation hardness, the fast response and the compact detector dimensions~\cite{mavro}
very well adapted for the experimental conditions encountered in the very forward region
at the LHC. A detailed description of the operation principle and, in particular, of the 
light-guide performances have been provided in reference~\cite{castor_protoI}. In 
section~\ref{sec:WQ_plates} we describe the active (quartz) and passive (tungsten) 
materials of the calorimeter considered in this second beam test. Section~\ref{sec:light_read} 
discusses the characteristics of the two types of photodetectors (photomultipliers and 
avalanche photodiodes) tested.


\subsection{Tungsten - Quartz plates}
\label{sec:WQ_plates}

\begin{figure*}[htbp] 
\begin{center}
\includegraphics[width=12cm]{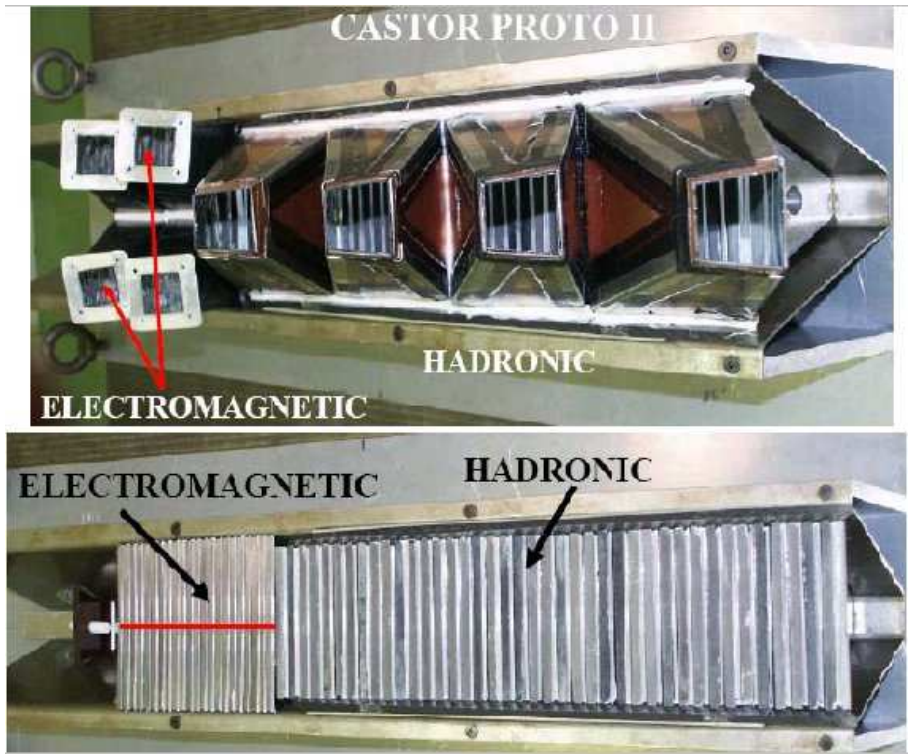}
\caption{Upper photograph of the W/Q-plates of the CASTOR prototype-II 
showing the EM and HAD sections (lower picture) and the light guides (upper picture) 
in the semi-octant (octant) geometry of the EM (HAD) sections respectively.}
\label{fig:WQ_plates_pic}
\end{center}
\end{figure*}

The calorimeter is constructed from layers of tungsten (W: $\lambda_I$= 10.0 cm, $X_0$= 0.365 cm, 
density= 18.5 g/cm$^3$) plates as absorber and fused silica quartz (Q) plates as active medium 
(see Fig.~\ref{fig:WQ_plates_pic}). For the electromagnetic section, the 
W-plates have a thickness of 3 mm and the Q-plates 1.5 mm. For the hadronic section, 
the W- and Q-plates have a larger thicknesses of 5 mm and 2 mm, respectively. 
The W/Q-plates are inclined 45$^\circ$ with respect to the direction of the impinging 
particles, in order to maximize the \v{C}erenkov light output in the quartz. 
Each individual combination of W/Q-plates is called a sampling unit (SU).
The large sides of the Q plates were covered with Tyvek paper, to protect them
from damage by the tungsten plates and also to diffuse back the escaping light. 
The perimeter sides -- except the top one -- were painted with white reflecting paint. 
The top edge of the W plates had just a machined finish.

In the EM section, each sampling unit (SU) corresponds to 1.218 $X_0$, or 
4.88$\times$10$^{-2}$ $\lambda_I$. Each readout unit (RU) consists of 11 SUs and is 
13.4 $X_0$, or 0.536 $\lambda_I$ deep. The EM section is divided in two successive 
RUs and has a total length of 26.8 $X_0$ and 1.072$\lambda_I$ lengths. In the hadronic section, 
a sampling unit corresponds to 7.96$\cdot 10^{-2}\,\lambda_I$. Each readout unit consists 
of 10 SUs and is 0.796 $\lambda_I$ deep. The HAD section has 4 RUs, corresponding to 3.186 $\lambda_I$.
 
In total, the whole prototype has 4.26 $\lambda_I$. For some runs with pions, we inserted 
an additional inactive absorber of 1.03 $\lambda_I$ in front of the calorimeter, in order to make 
the EM section act as a hadronic one, increasing the total depth of the prototype to 
5.3$\lambda_I$.


\subsection{Photodetectors}
\label{sec:light_read}

\begin{figure*}[htbp] 
\begin{center}
\includegraphics[width=12cm]{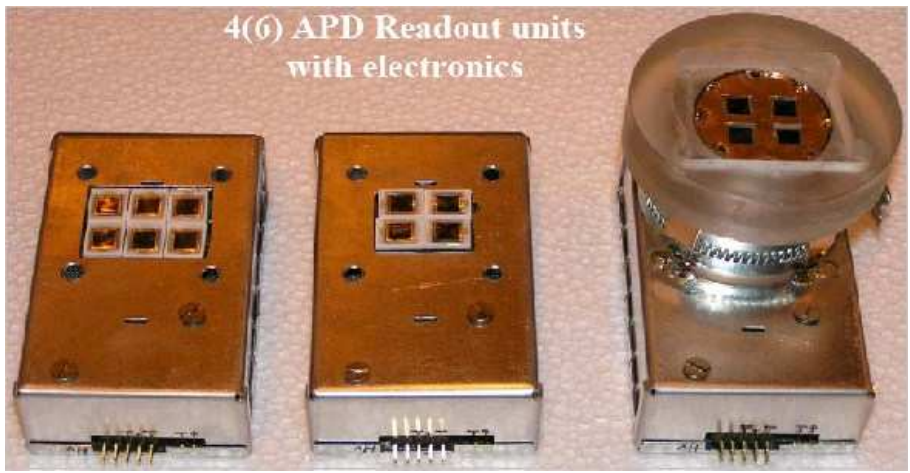}
\caption{Assembled APD readout units with 4 and 6 APDs.}
\label{fig:apd_pics}
\end{center}
\end{figure*}

The \v{C}erenkov light emitted by the quartz plates is collected and transmitted to 
photodetector devices through air-core light-guides. All light guides of Prototype-II 
were equipped with Dupont [AlO+ SiO$_2$+TiO$_2$] reflective foil with the same 
characteristics discussed in~\cite{castor_protoI}.
As photodetectors we used a matrix of 4 or 6 Hamamatsu S8148 APDs (developed 
originally for the CMS electromagnetic calorimeter~\cite{apd1}), as well as two 
different types of PMTs. The total area of the APDs was 1 cm$^2$ (for 4 APDs) 
and 1.5 cm$^2$ (for 6 APDs), see Fig.~\ref{fig:apd_pics}. The phototubes were 
positioned only on one side of the EM section of the prototype, for comparison 
with the APDs during the electron beam tests. The two types of PMTs used were 
respectively: (i) a Hamamatsu R7899 PMT, and (ii) a radiation-hard multi-mesh, 
small size PMT FEU-187 from RIE St. Petersburg, with cathode area 
$\sim$2 cm$^2$~\cite{castor_protoI}.

\section{Beam tests}
\label{sec:beam_tests}

\begin{figure*}[htbp] 
\begin{center}
\includegraphics[width=12cm]{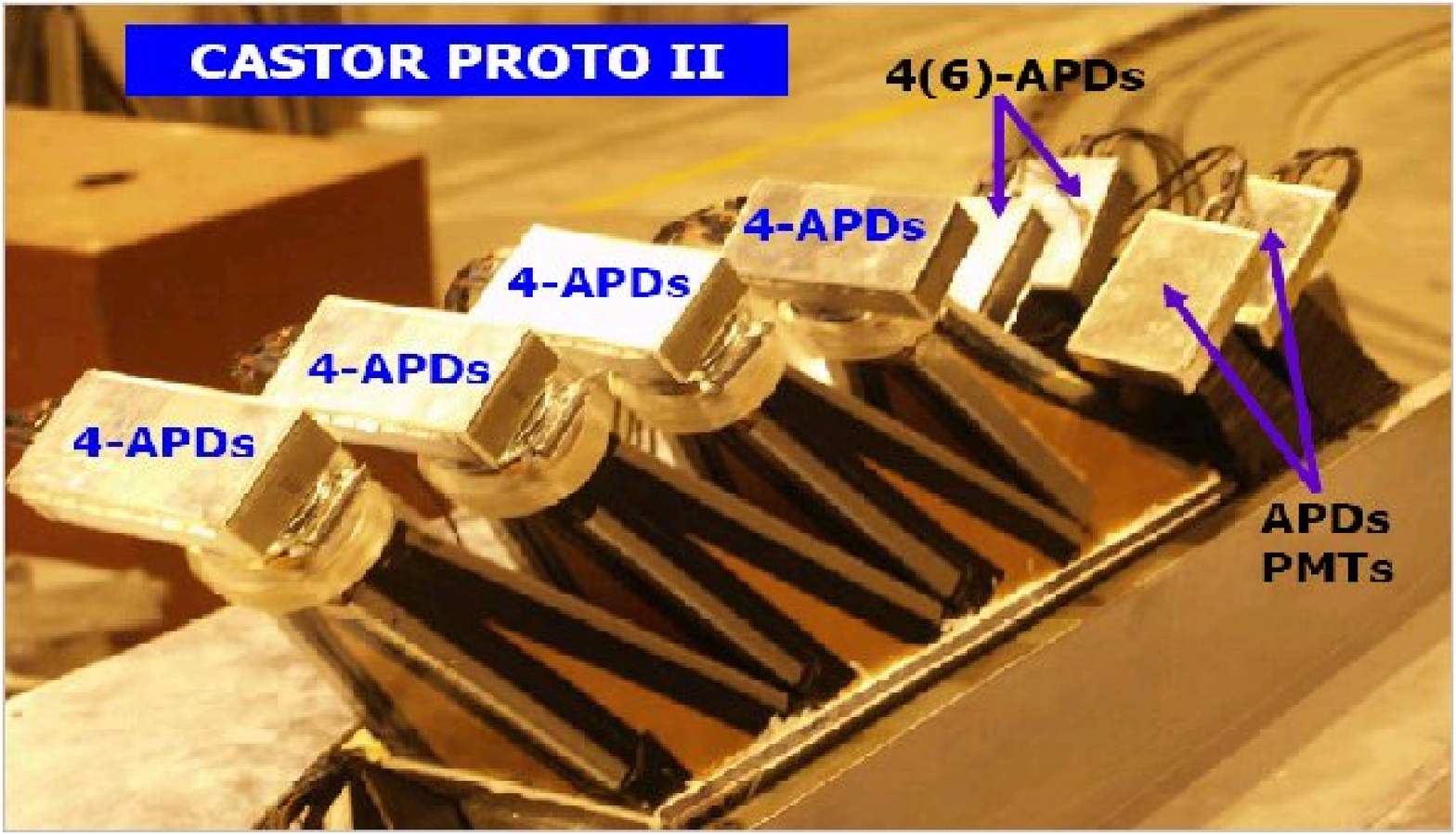}
\caption{Assembled prototype II on the moving table in the CERN/SPS H2 beam line. 
Only the APD readout units are shown.}
\label{fig:protoII_pic2}
\end{center}
\end{figure*}

The beam test of prototype II took place in October 2004 at the H2 beam line 
of the SPS at CERN. Electron, hadron ($\pi^-$) and muon ($\mu^-$) beams of several 
energies were used. The energy responses (linearity, resolution) of the 
electromagnetic and hadronic calorimeters were obtained with energy scans 
with: 20-200 GeV electrons, 20-350 GeV pions, as well as 50, 150 GeV muons.
The calorimeter prototype was placed on a platform movable with respect to 
the beam in both horizontal and vertical ($x,y$) directions
(see Figure~\ref{fig:protoII_pic2}).
A telescope of finger scintillator detectors and wire chambers were installed upstream 
of the prototype, giving precise information on the position of each particle 
hitting the calorimeter. In this way, we were able to know the beam profile 
and also select particular regions of the beam profile for the spatial 
resolution analyses. We note that the typical visible transverse sizes of hadronic and 
electromagnetic showers in quartz calorimeters are $\cal{O}$(5-10 cm), $\cal{O}$(10 mm) 
resp. (for 95\% signal containement), i.e. are a factor 3 to 4 times narrower than
those in ``standard'' (scintillation) calorimeters~\cite{mavro}.

\begin{figure*}[htbp] 
\begin{center}
\includegraphics[width=12cm,height=8cm]{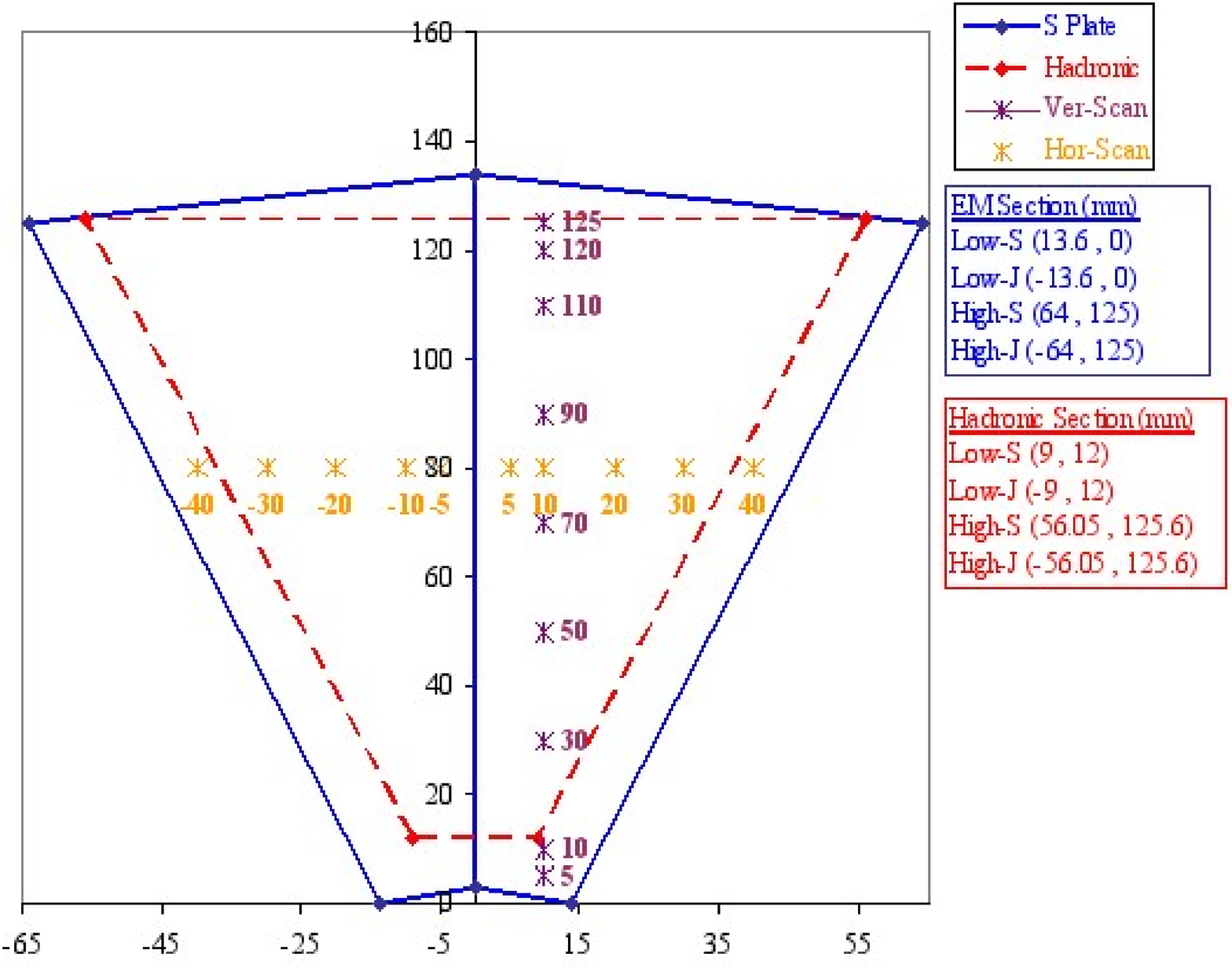}
\caption{Projection of the EM (blue) and HAD (red) sections onto a 45$^\circ$ plane. 
The numbers indicate the $x-y$ coordinates of the beam impact points 
(indicated by the '$\star$' symbol) used in the horizontal and vertical scans.}
\label{fig:castor_projection}
\end{center}
\end{figure*}

Figure~\ref{fig:castor_projection} shows the two semi-octants of the electromagnetic 
(blue) and the octant of the hadronic (red) sections, as seen projected onto a plane 
at 45$^\circ$ with respect to the beam axis. We notice that there is no complete overlap 
of the two sections, due to the different sizes of the W/Q-plates available. 
The horizontal and vertical numbers correspond to distances along the plate 
($x-y$ coordinates) of the points used for the horizontal and vertical scans.

Table~\ref{tab:xy_coords} lists the ($x,y$) coordinates of the impact points of the 
horizontal and vertical scans for both electron and hadron beams. The location of these 
points on the 45$^\circ$ projection of the semi-octant sectors is shown in 
Figure~\ref{fig:castor_projection}. The beam profile for each point was subdivided 
into a number of smaller parts, each of diameter $\sim$1-2 mm, so that we obtained 
more impact points in total.

\begin{table*}[htbp]
\begin{center}
\caption{The ($x,y$) coordinates (mm) of the impact points of the horizontal 
and vertical scans for both electron and hadron beams.}
\vskip 0.2cm
\label{tab:xy_coords}
\renewcommand{\arraystretch}{1.5}
\begin{tabular}{|ccc|ccc|} \hline\hline
Electron SCAN &  &  &  &  & \\ \hline  
Vertical Scan & $x$ & $y$ & Horizontal Scan & $x$ & $y$ \\  
A & 10 & 5  &  A'  &  -40  &  80 \\  
B & 10 & 10  &  B'  &  -30  &  80 \\  
C & 10 & 30  &  C'  &  -20  &  80 \\  
D & 10 & 50  &  D'  &  -10  &  80 \\  
E & 10 & 70  &  E'  &  -5  &  80 \\  
F & 10 & 90  &  F'  &  5  &  80 \\  
G & 10 & 110  &  G'  &  10  &  80 \\  
H & 10 & 120  &  H'  &  20  &  80 \\  
I & 10 & 125  &  I'  &  30  &  80 \\  
  &    &    &  J'  &  40  &  80 \\  \hline\hline

Hadron SCAN  &    &    &    &   &  \\ \hline
Vertical Scan & $x$ & $y$  &  Horizontal Scan & $x$  &  $y$  \\  
A & 10 & 30  &  A' & -30  &  80 \\  
B & 10 & 50  &  B' & -20  &  80 \\  
C & 10 & 70  &  C' & -10  &  80 \\  
D & 10 & 90  &  D' & 0  &  80  \\  
E & 10 & 110  &  E' & 10  &  80  \\  
F & 10 & 120  &  F' & 20  &  80  \\  
  &    &    &  G' & 30  &  80  \\  \hline \hline
\end{tabular}
\renewcommand{\arraystretch}{1}
\end{center}
\end{table*}

\section{Electron beam tests}
\label{sec:e_beam}

Electron beams of energy 20-200 GeV were used to test the energy 
linearity and resolution as well as the position resolution of the EM 
section of the prototype. 


\subsection{Energy response}
\label{sec:en_response}

\begin{figure*}[htbp] 
\begin{center}
\includegraphics[width=11cm,height=16cm,angle=-90]{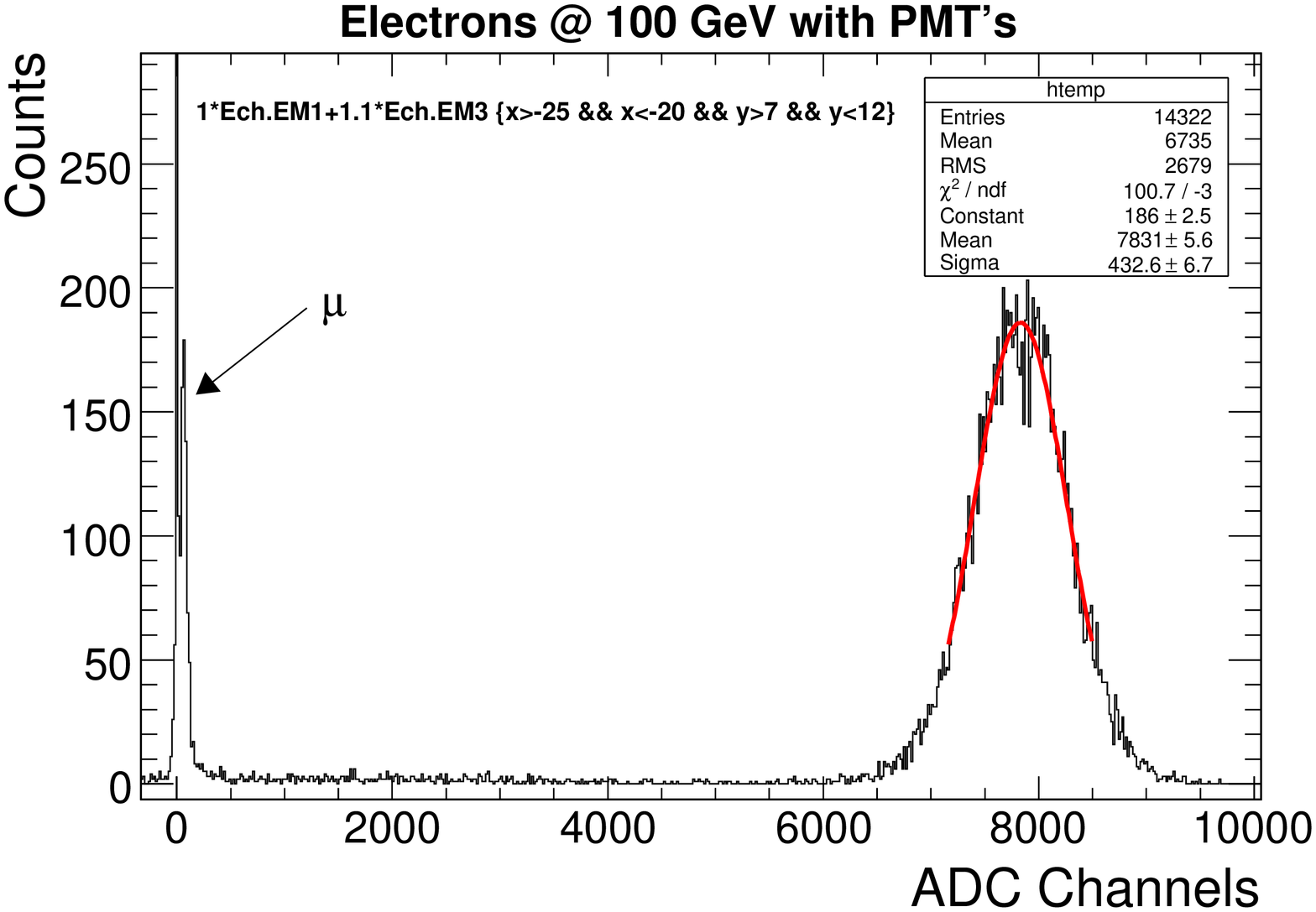}
\caption{Energy response of the EM calorimeter equipped with PMTs 
to 100 GeV electrons (and residual beam muons).}
\label{fig:em_spectrum}
\end{center}
\end{figure*}

\begin{figure*}[htbp] 
\begin{center}
\includegraphics[width=5.9cm,angle=-90]{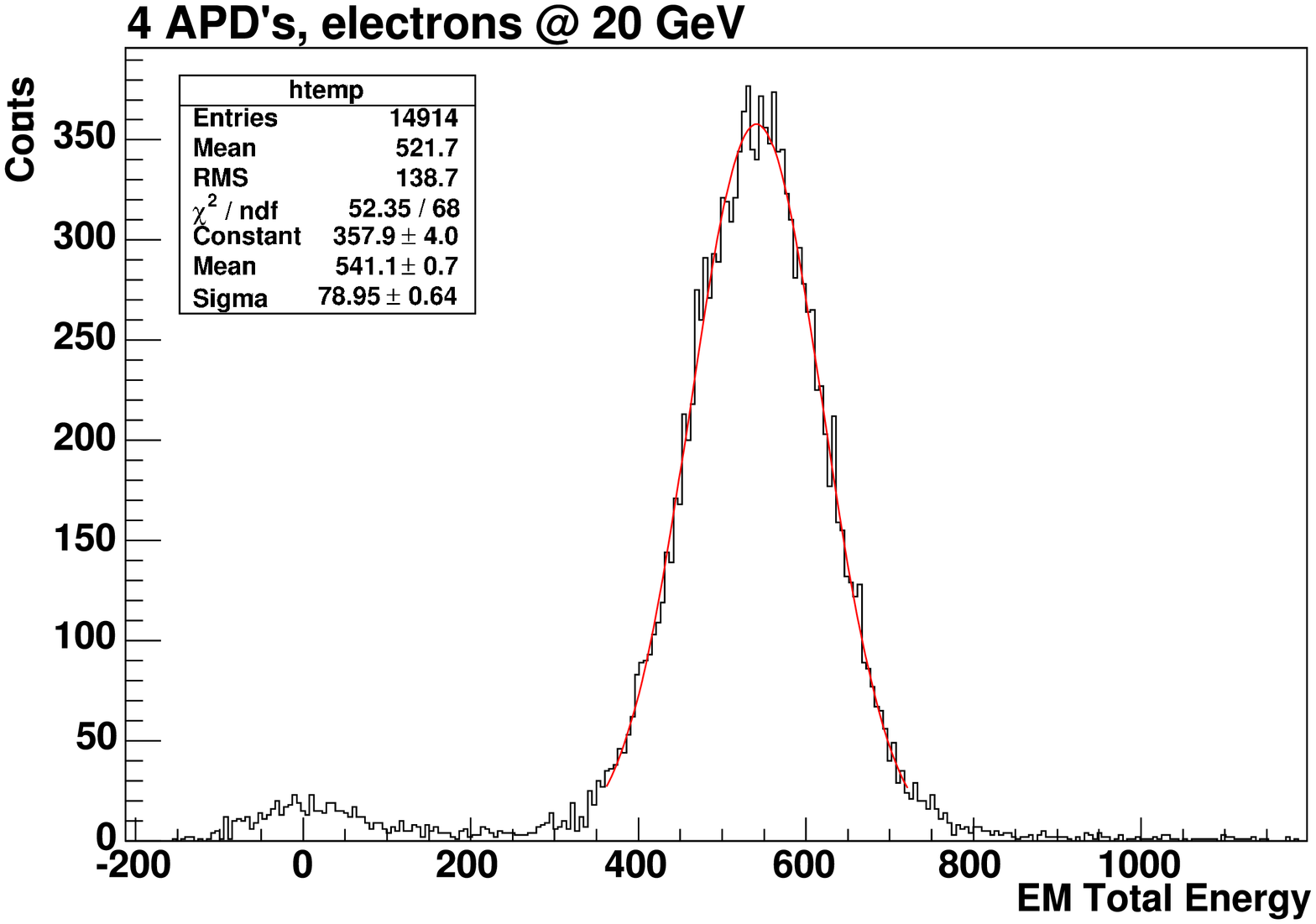}
\includegraphics[width=5.9cm,angle=-90]{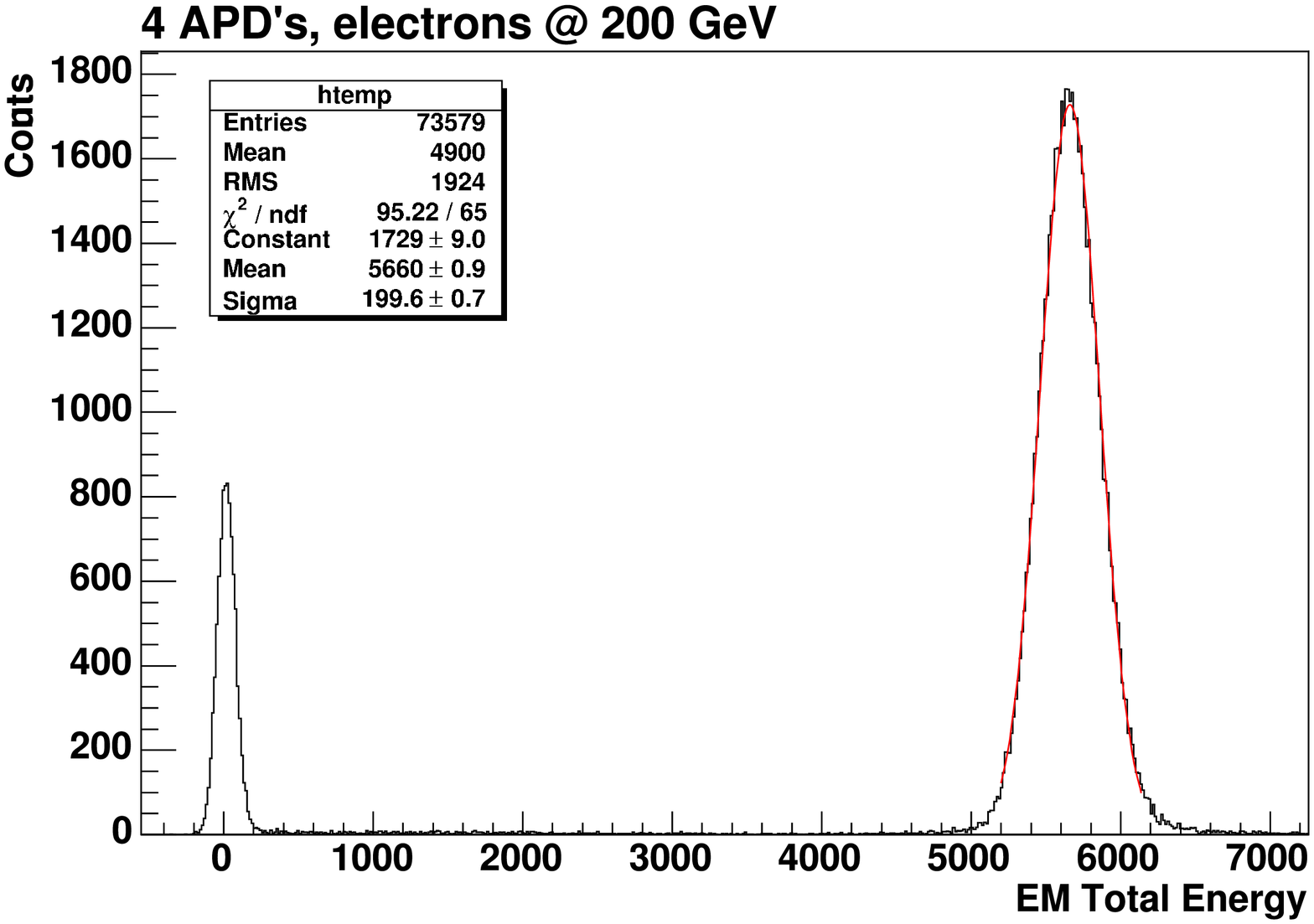}
\includegraphics[width=5.9cm,angle=-90]{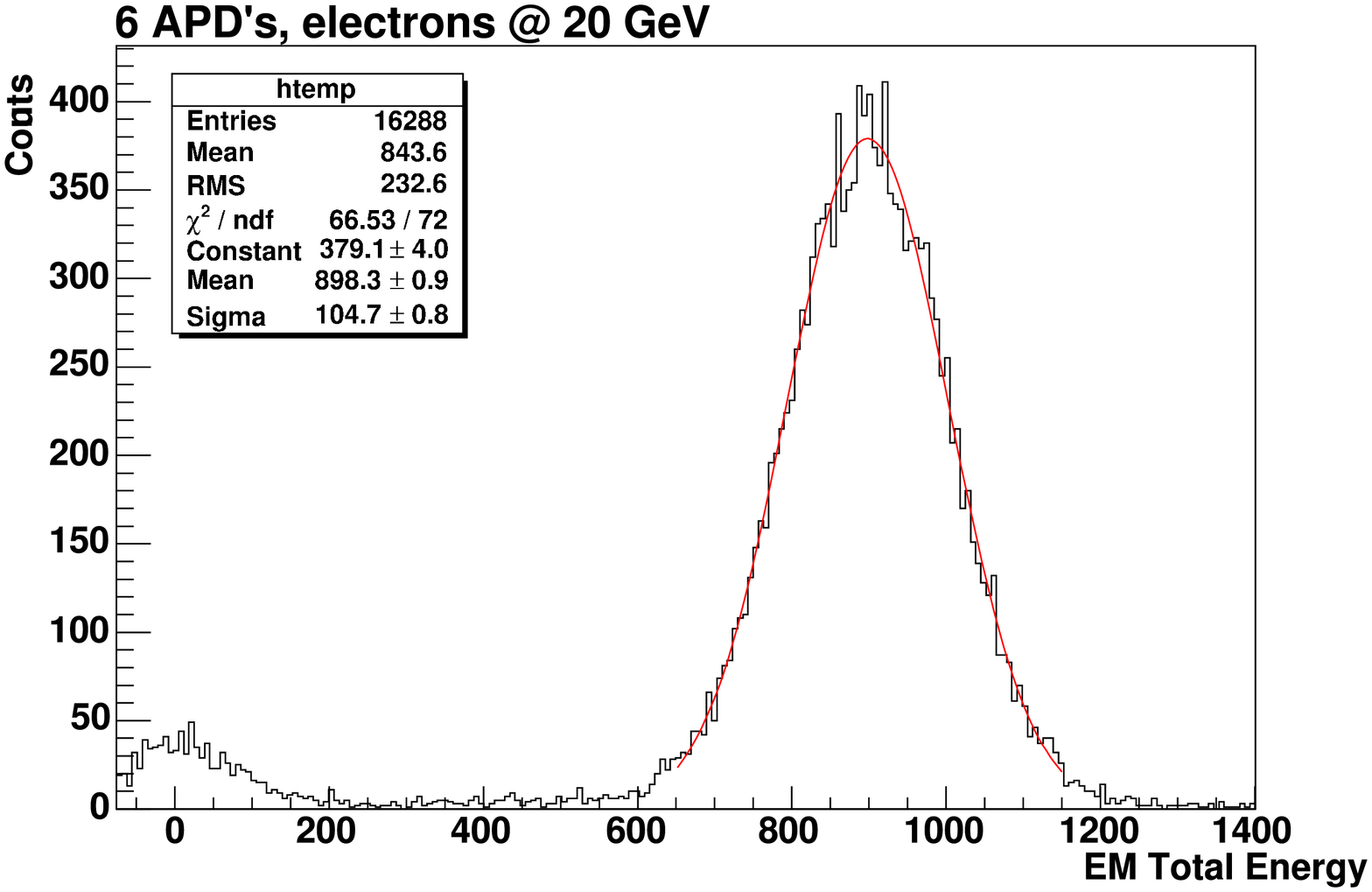}
\includegraphics[width=5.9cm,angle=-90]{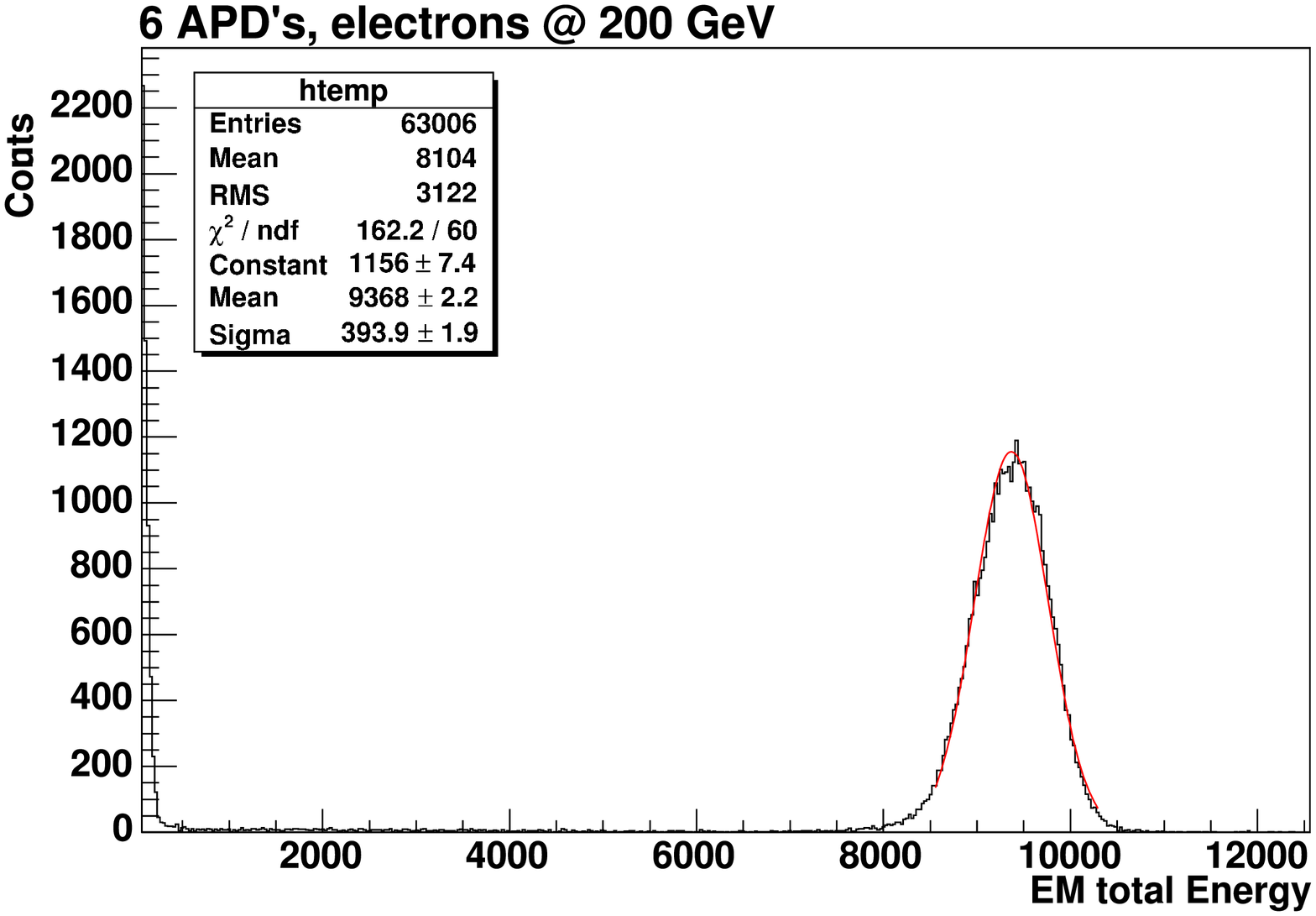}
\caption{Energy response of the EM calorimeter to electron beams of 
20 and 200 GeV obtained with 4 APDs (upper plots) and 6 APDs (bottom plots).}
\label{fig:em_spectrum2}
\end{center}
\end{figure*}

A typical spectrum measured with 100 GeV electrons incident on the EM section 
of the prototype, equipped with PMTs, is shown in Figure~\ref{fig:em_spectrum}. 
Residual muons in the electron beam are also seen as minimum ionizing 
particle (MIPs) just above the pedestal. The energy response of the 
calorimeter is found to be Gaussian for all energies. Figure~\ref{fig:em_spectrum2} 
shows the energy response for 20 and 200 GeV electron beams, obtained with 
4 and 6 APDs respectively.

\begin{figure*}[htbp] 
\begin{center}
\includegraphics[width=9cm,angle=-90]{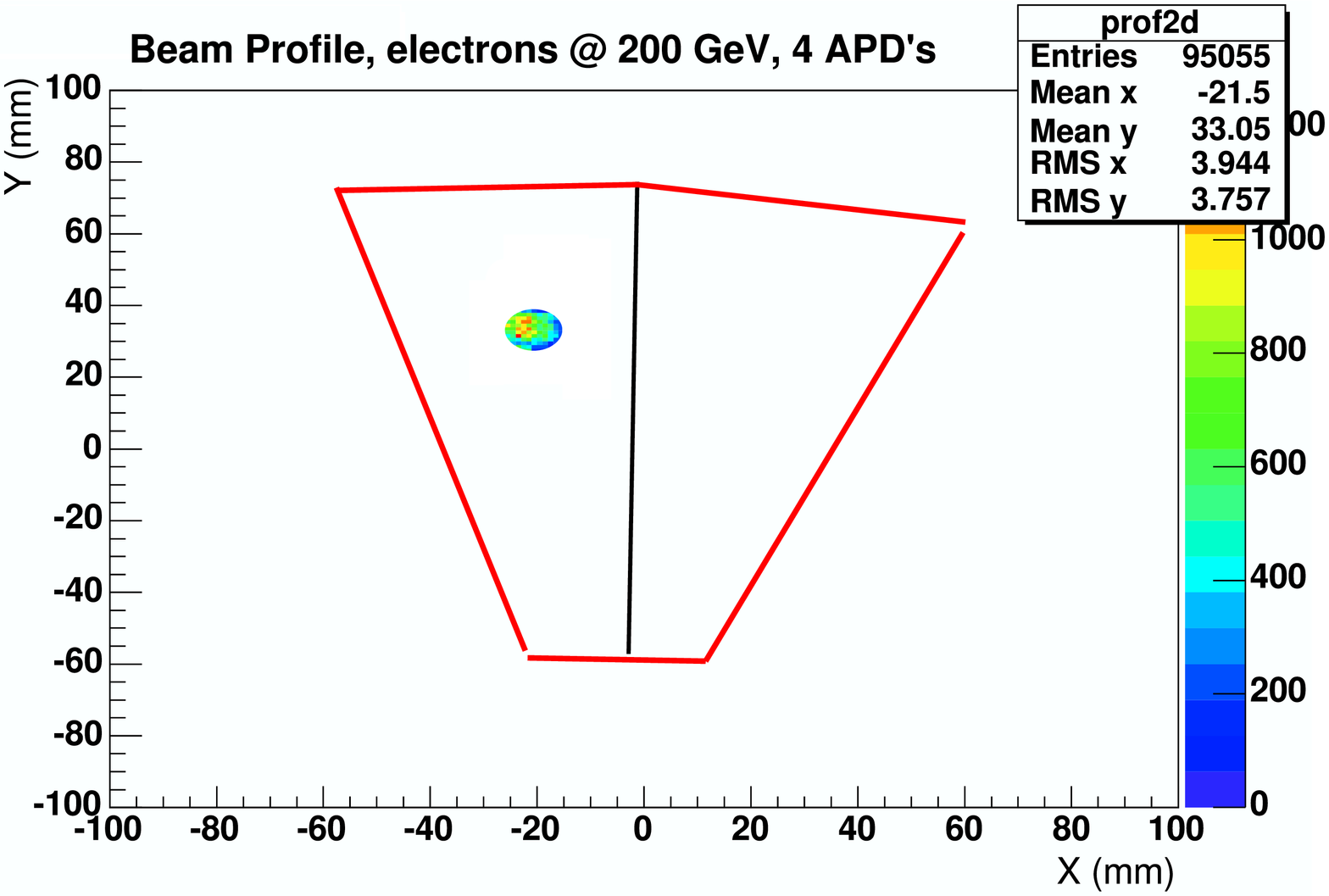}
\caption{Profile of 200 GeV electron impinging on the left semi-octant of   
the calorimeter, as measured by the scintillator-wire-chamber telescope 
upstream of the prototype.}
\label{fig:elec_impact}
\end{center}
\end{figure*}

\begin{figure*}[htbp] 
\begin{center}
\vskip 0.8cm
\hskip -0.8cm
\includegraphics[width=8cm,angle=-90]{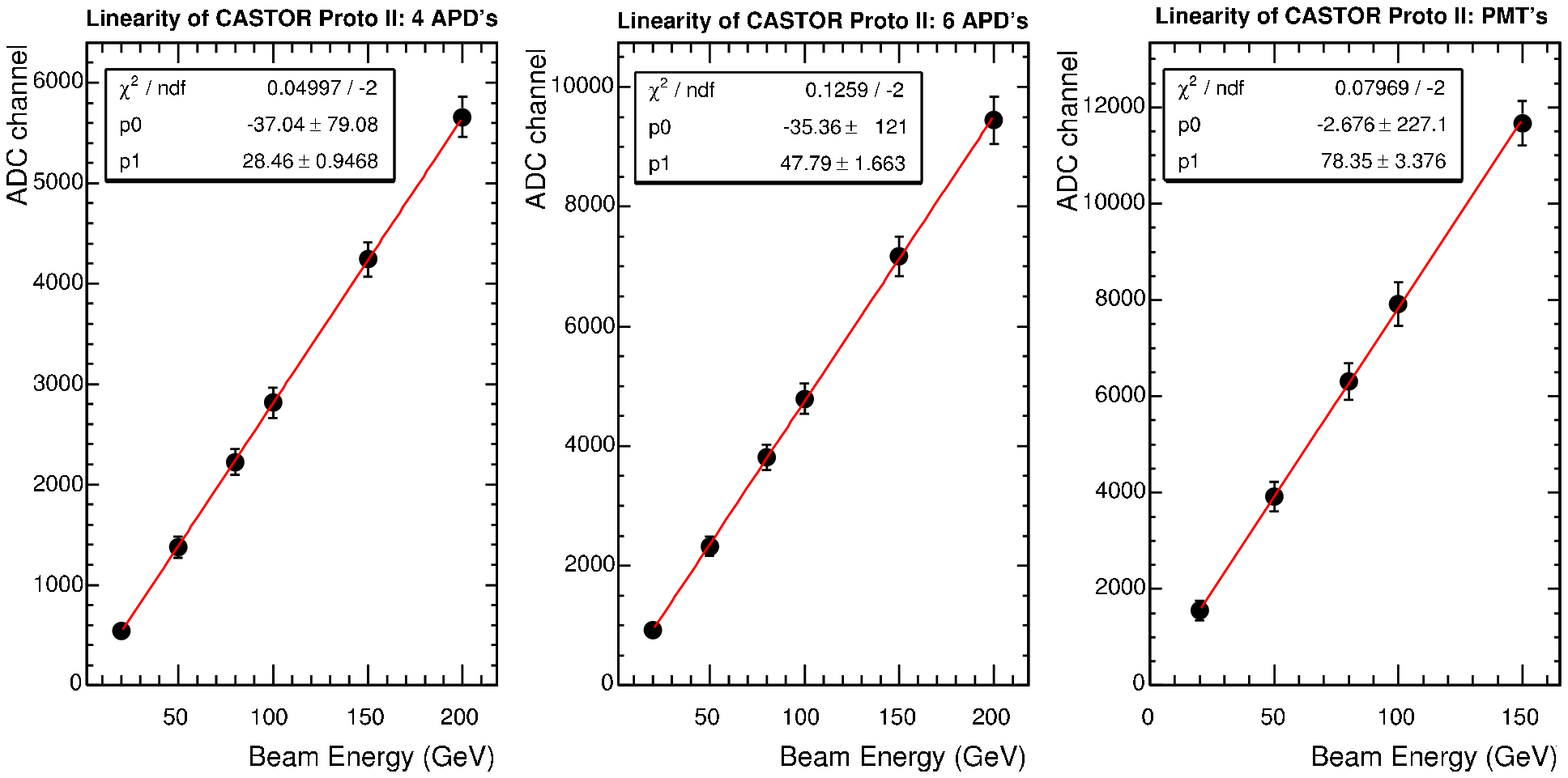}
\caption{Energy response linearity (signal peak-position versus beam energy) 
of the EM section, obtained with different photodetectors: 
4 APDs (left), 6 APDs (center), and PMTs (right).}
\label{fig:em_linearity}
\end{center}
\end{figure*}

\noindent {\bf Energy Linearity: }
To study the linearity of the EM calorimeter response as a function of 
electron-beam energy, a central point (Fig.~\ref{fig:elec_impact}) in the 
two different azimuthal sectors has been exposed to beams of various energies. 
The distributions of signal amplitudes, after introducing the cuts on the 
spatial profile of the beam (a circle of radius 2 mm), are in most cases symmetric 
and well fitted by a Gaussian function. The peak signal position, obtained 
for the three photodetector configurations, is plotted as a function of the 
beam energy in Figure~\ref{fig:em_linearity}.

For all configurations, the calorimeter response is found to be linear in the energy 
range explored. The average signal amplitude, expressed in units of ADC channels, 
is satisfactorily fitted by the formula:

\begin{eqnarray}
ADC & = & a + b \times E 
\end{eqnarray}

where the energy E is in GeV. The fitted values of the parameters for each configuration 
are shown in the insets of each plot in Fig.~\ref{fig:em_linearity}.\\ 


\noindent {\bf Energy Resolution: }
\noindent The relative energy resolution of the calorimeter has been studied by plotting the 
normalized width of the Gaussian signal amplitudes, $\sigma/E$, with respect to the incident 
beam electron energy, $E$(GeV) and fitting the data points with two different functional 
forms~\cite{castor_protoI}: 

\begin{eqnarray}
\sigma/E & = & p_0 + p_1/\sqrt{E} 			\label{eq:2} \\
\sigma/E & = & p_0 \oplus p_1/\sqrt{E} \oplus p_2/E 	\label{eq:3}
\end{eqnarray}
 
where the $\oplus$ indicates that the terms are added in quadrature. 
In principle, three general terms contribute to the energy resolution in calorimeters:
\begin{enumerate}
\item The constant term, $p_0$, related to imperfections of the calorimetry,
signal generation and collection non-uniformity, calibration errors and fluctuations
in the energy leakage, which limit the resolution at high energies. 
\item The stochastic or sampling term, $p_1$, due to intrinsic shower photon 
statistics, characterizes the fluctuations in the signal generating process.
\item The noise term, $p_2$, includes the electronic noise contribution from 
capacitance and dark current which (due to its steep $1/E$ dependence) is only 
important for low energies.
\end{enumerate}

\begin{figure*}[htbp] 
\begin{center}
\includegraphics[width=10cm,angle=-90]{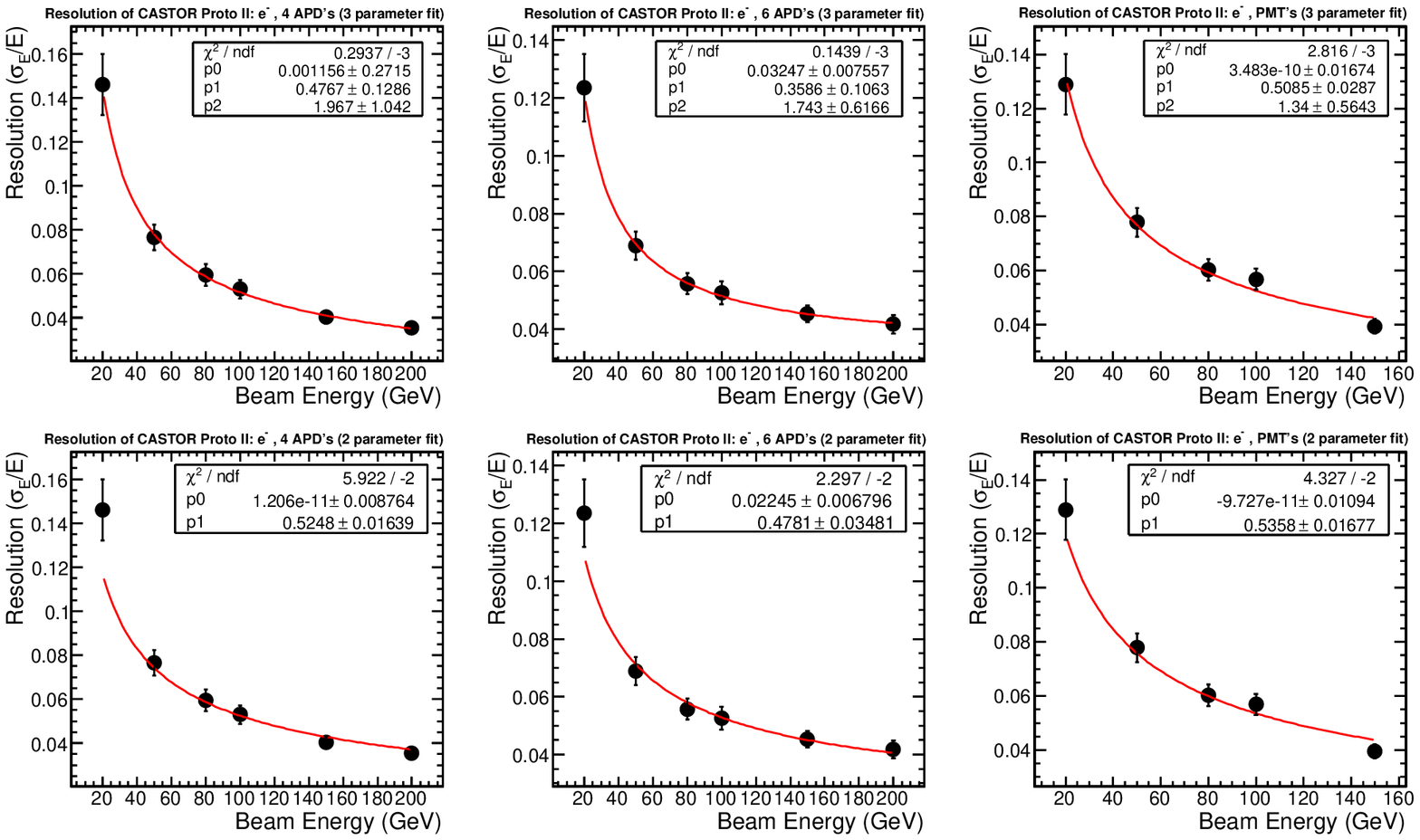}
\caption{Energy resolution (signal peak width versus beam energy) of the 
prototype EM section, obtained with the three readout configurations considered: 
4 APDs (left), 6 APDs (center), and PMTs (right): 3-parameters fit 
Eq.~(\ref{eq:3}) (top); 2-parameters fit Eq.~(\ref{eq:2}) (bottom).}
\label{fig:en_fit}
\end{center}
\end{figure*}

Figure~\ref{fig:en_fit} shows the fit to the data with expressions (\ref{eq:2}) and 
(\ref{eq:3}). Both parametrizations satisfactorily fit the data. In Table~\ref{tab:en_fit_params} 
we summarize the fit parameters for both parameterizations and the three readout configurations. 
The measured stochastic term $p_1$ is in the range 36\% - 54\%. We notice too that 
the constant term $p_0$ is close to zero for all options. It should be noted that though 
the APDs are very sensitive to both voltage and temperature changes, there was no 
stabilization used for this test.\\

\begin{table*}[htbp]
\begin{center}
\caption{Energy resolution parameters of the EM calorimeter prototype as obtained 
from the measured electron beam energy resolution and Eqs. (\ref{eq:2}), (\ref{eq:3}).}
\vskip 0.2cm
\label{tab:en_fit_params}
\renewcommand{\arraystretch}{1.5}
\begin{tabular}{|c|c|c|c|c|c|} \hline\hline
Photodetector(s) & Fit function & $p_0$  & $p_1$ (GeV$^{1/2}$) & $p_2$ (GeV) & $\chi^2/$ndf \\ \hline  
 4 APDs & (2) & 1.2e-11 $\pm$ 8.7e-3 & 0.525 $\pm$ 0.0163  & -  &  5.92/4 \\  
 4 APDs & (3) & 1.1e-3 $\pm$ 0.21 & 0.477 $\pm$ 9.65e-2  &  1.97 $\pm$ 0.70  &  0.29/3 \\  \hline
 6 APDs & (2) & 2.24e-2 $\pm$ 6.80e-3 & 0.478 $\pm$ 0.0348  & -  &  2.30/4 \\  
 6 APDs & (3) & 3.25e-2 $\pm$ 7.56e-2 & 0.358 $\pm$ 0.106  &  1.74 $\pm$ 0.62  &  0.14/3 \\  \hline
 PMTs   & (2) & 9.7e-11 $\pm$ 1.1e-2 & 0.536 $\pm$ 0.0168  & -  &  4.33/3 \\  
 PMTs   & (3) & 3.5e-10 $\pm$ 1.7e-2 & 0.508 $\pm$ 0.029  &  1.34 $\pm$ 0.56  &  2.82/2 \\  \hline
\end{tabular}
\renewcommand{\arraystretch}{1}
\end{center}
\end{table*}

%

\noindent {\bf Spatial Response: }
The purpose of the area scanning was to check the uniformity of the EM calorimeter 
response to electrons hitting at different points on the sector area, as well as to 
assess the amount of edge effects and lateral leakage from the calorimeter, which could lead 
to cross-talk between neighboring sectors. Figure~\ref{fig:elec_impact} shows the typical 
profile of the electron beam hitting the left semi-octant of the prototype. The width 
of the EM shower and the percentage of the containment close to the edge were estimated by 
varying the horizontal and vertical hit positions of the incident beam according to the ($x,y$) 
coordinates shown in Fig.~\ref{fig:elec_impact} and listed in Table~\ref{tab:xy_coords}.

\begin{figure*}[htbp] 
\begin{center}
\hskip -0.8cm
\includegraphics[width=9cm,angle=-90]{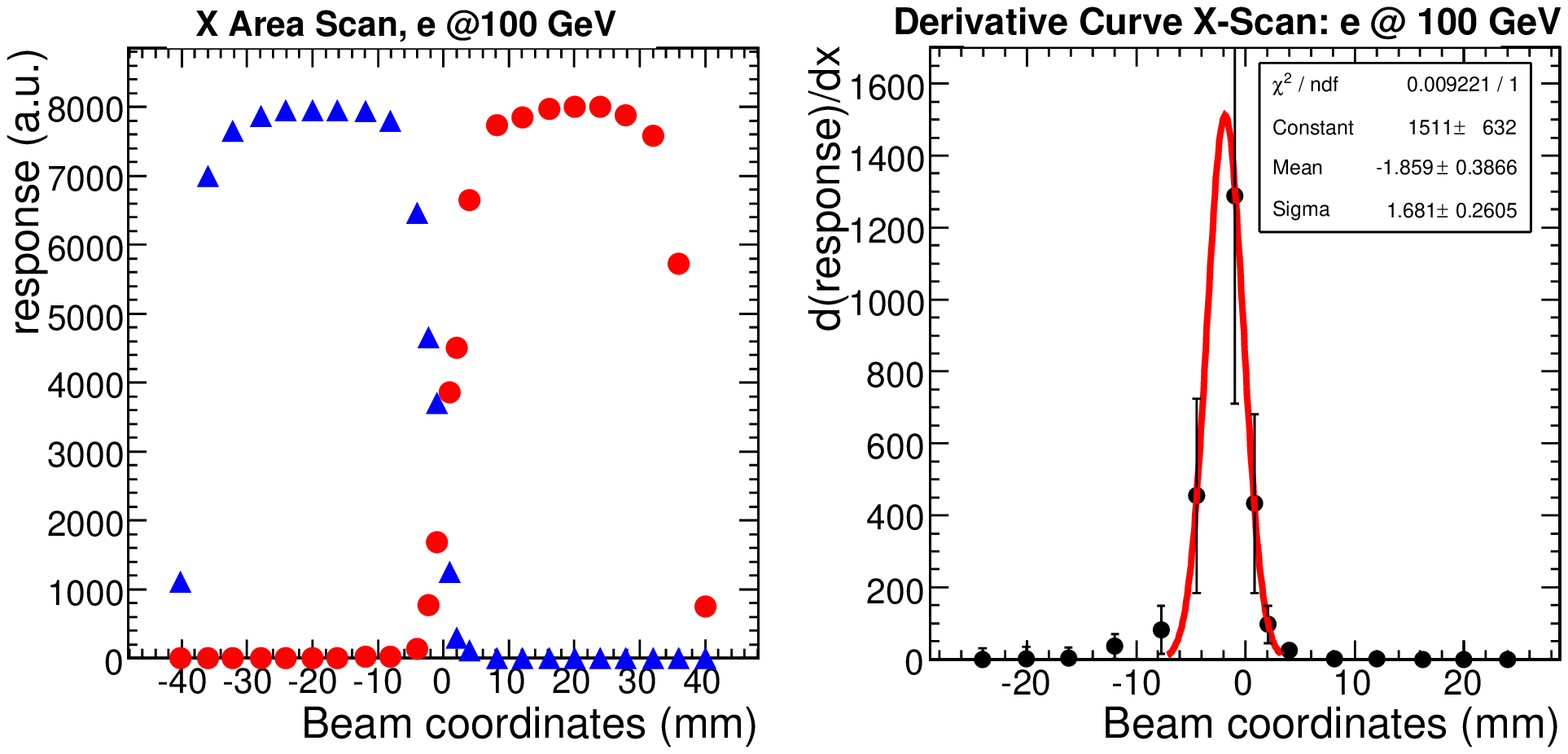}
\caption{ (a) Response of the left and right semi-octant sectors of the EM section as 
the beam scans the front face of the calorimeter. (b) The derivative of the response with 
respect to $x$, indicating the width of the EM shower.}
\label{fig:em_esp_response}
\end{center}
\end{figure*}

The results of the horizontal-scan analysis are shown in Figure~\ref{fig:em_esp_response} 
for the 4 APDs readout configuration. Figure~\ref{fig:em_esp_response}(a) shows the response 
of the two adjacent (left-right) EM semi-octants as the beam impact point moves across 
the front face of the calorimeter. The sigmoid nature of each response curve is evident. 
In Figure~\ref{fig:em_esp_response}(b), the $x-$derivative of the response is calculated, 
giving the width of the electromagnetic shower. We observe that one standard deviation 
amounts to 1.7 mm.

\begin{figure*}[htbp] 
\begin{center}
\includegraphics[width=14cm]{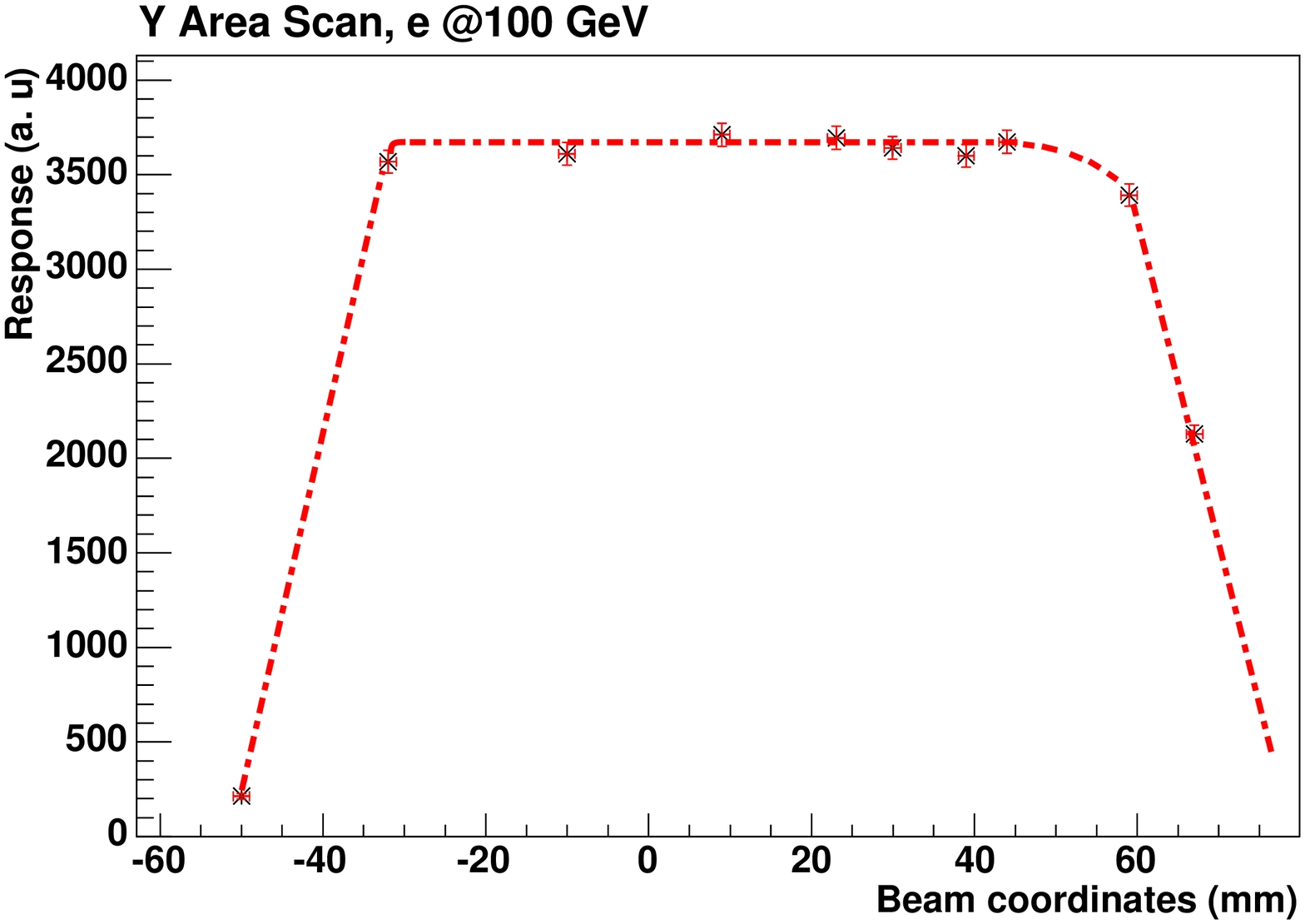}
\caption{Spatial scan along the $y$-direction of one EM sector. The impact points are those listed in 
Table~\protect\ref{tab:xy_coords} and shown in Fig.~\protect\ref{fig:castor_projection} [Note that,
at variance with the beam-coordinate-system used in Fig.~\protect\ref{fig:castor_projection}, we use 
here a calorimeter-coordinate-system, based on the position of the calorimeter on the moving platform].}
\label{fig:em_y_scan}
\end{center}
\end{figure*}

The vertical-scan covered the entire height of the semi-octant EM sector, with impact 
points shown in Figure~\ref{fig:castor_projection} and listed in Table~\ref{tab:xy_coords}. 
The results of this scan are shown in Figure~\ref{fig:em_y_scan}. We notice the 
abrupt fall at the lower end of the sector past the point "A" and the more gradual fall 
at the upper end, the later due to the shower particles directly hitting the light guide.

\subsection{Pion beam tests}

Pions of energy 20--350 GeV were used for the study of the hadronic 
energy and position responses of the CASTOR prototype II. In order to 
increase the interaction depth of the calorimeter, an inactive absorber 
of 1.03$\lambda_I$ was inserted in front of the EM calorimeter, increasing 
the total depth to 5.3$\lambda_I$. This had also as a result to make the two 
first (EM) RUs effectively act, in depth, as part of the hadronic section.\\

\begin{figure*}[htbp] 
\begin{center}
\includegraphics[width=10cm,angle=-90]{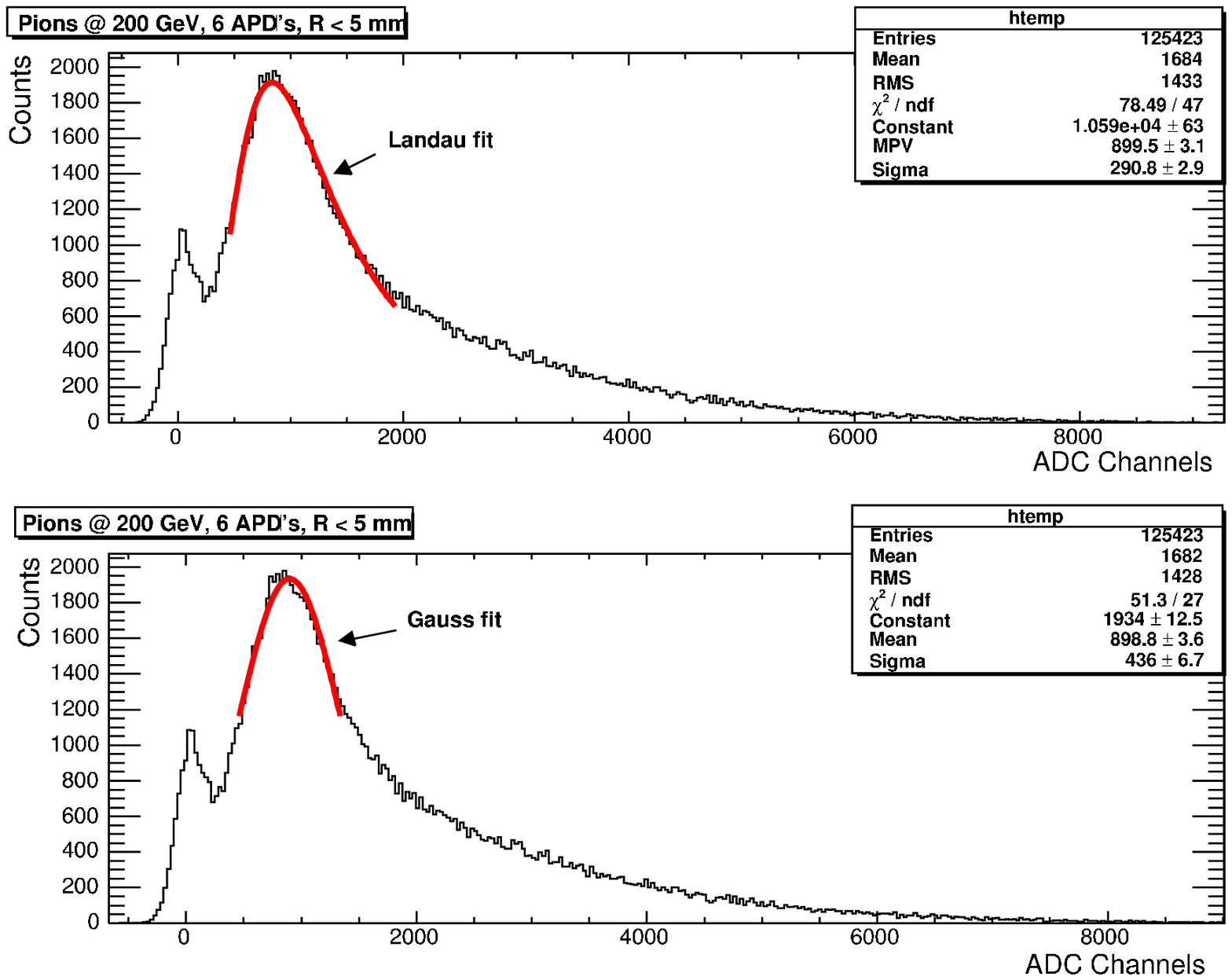}
\caption{Total energy spectra (ADC channel counts) measured in the prototype-II
for the pion beam of 200 GeV and 6 APDs in the EM section. The pion peak is 
fitted to a Landau (top plot) and Gaussian (bottom plot) curve with fit parameters
reported in the inset. The peak to the left is the pedestal.}
\label{fig:had_spectrum}
\end{center}
\end{figure*}

\noindent {\bf Energy Response: }
Typical spectra, obtained with 200 GeV pions incident on the prototype, 
are shown in Figure~\ref{fig:had_spectrum} where the distribution of 
the total energy measured in both (EM and HAD) parts of the calorimeter is plotted. 
During the different tests, the electromagnetic sections were equipped with 
4 or 6 APDs and the hadronic ones had 4 APDs in its readout units for all runs.
The total depth of the prototype (5.3$\lambda_I$) was not enough to contain 
the showers produced by the pion beams. We see that there is a long tail at
high energies indicating the leakage of energy from the back of the calorimeter.
However, no quantitative measurements of the leakage fraction were done at this stage.
The peak of the total pion energy measured by the prototype was fitted with 
a Gaussian and a Landau curve. The fitting ranges correspond roughly to
1- (2-)$\sigma$ around the peak for the Gaussian (Landau) distributions. 
We observe that the Landau parametrization fits the distribution better than 
the Gaussian one.

The energy response (position and width of the pion peak) was obtained by 
fitting both Gaussian and Landau curves to the spectrum measured for all beam 
energies. The corresponding hadronic energy linearity and resolution were thus 
obtained.\\

\begin{figure*}[htbp] 
\begin{center}
\includegraphics[width=12cm,angle=-90]{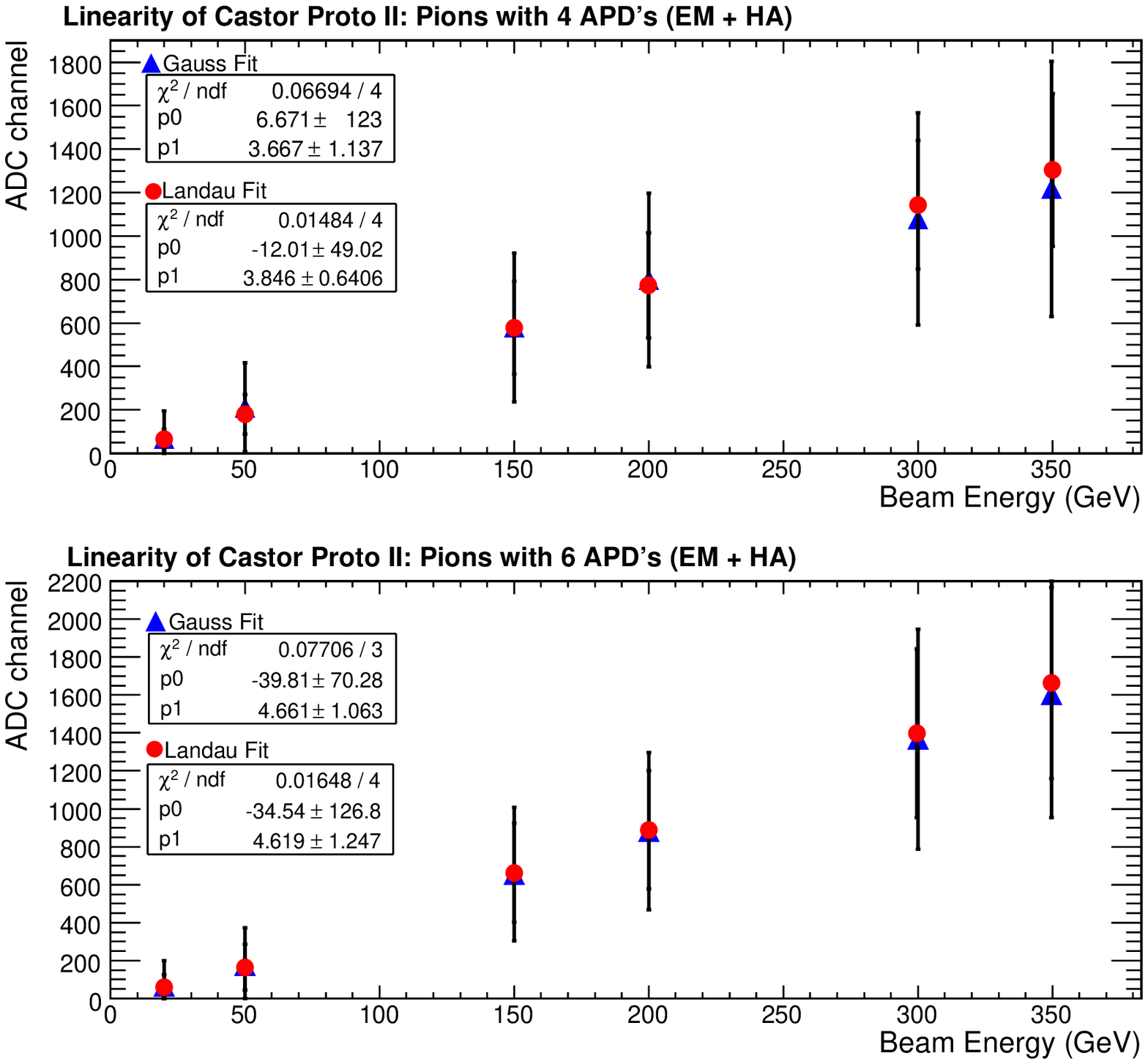}
\caption{Energy response linearity of the prototype calorimeter to pions 
of several energies, fitted to Gaussian (blue) and Landau (red) parametrizations.
The top (bottom) plot is obtained with 4 (6) APDs readout in the EM section.}
\label{fig:had_linearity}
\end{center}
\end{figure*}

\noindent {\bf Energy Linearity: }
Figure~\ref{fig:had_linearity} shows the linearity of the CASTOR prototype 
to incident pions as obtained by measuring the total energy deposited in the calorimeter 
sections and correlating the position of the pion peak with each corresponding beam 
energy. At higher energies, the Landau fit gives higher response, as expected, 
and an overall smaller statistical error.
 
\begin{figure*}[htbp] 
\begin{center}
\includegraphics[width=14cm]{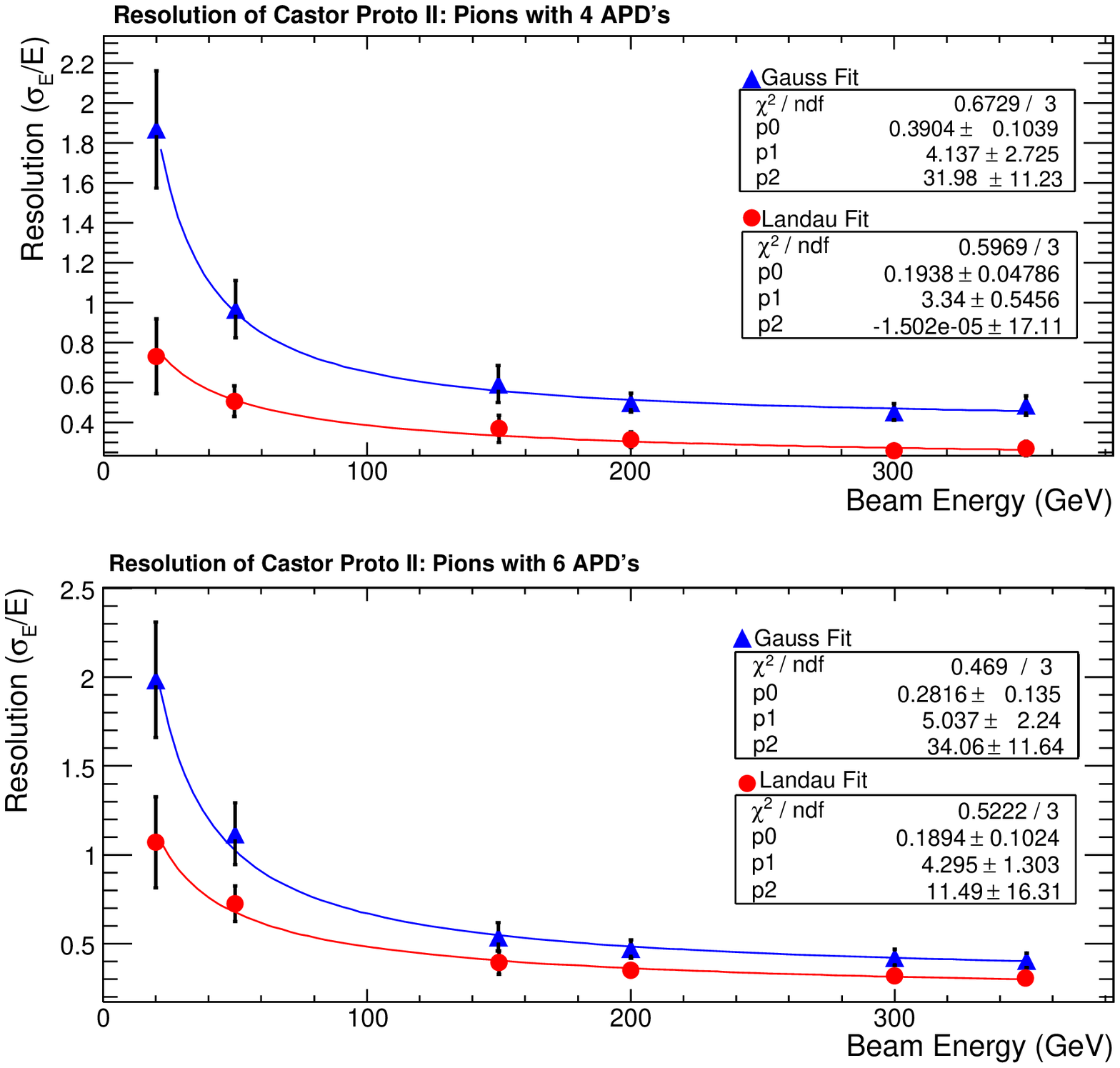}
\caption{Energy resolution, $\sigma/E$, of CASTOR prototype II to pion beams of several energies
obtained with 4 (top) and 6 (bottom) APDs readout in the EM section. The different fit parameters shown 
in the inset are obtained with Eq.~(\ref{eq:3}) and Eq.~(\ref{eq:2}) when the widths $\sigma$ 
of the pion peaks are fitted to a Gaussian or Landau distribution.}
\label{fig:had_en_response}
\end{center}
\end{figure*}

\noindent {\bf Energy Resolution: }
The relative energy resolution of the calo-rimeter has been studied by fitting 
the normalized width of the fitted signal amplitudes (peaks in Fig.~\ref{fig:had_spectrum}), 
$\sigma/E$, with respect to the incident pion beam energy, $E$(GeV), with the two 
functional forms (\ref{eq:2}) and (\ref{eq:3}). Figure~\ref{fig:had_en_response}
shows the obtained energy resolution of the prototype for pions of energy up to 
350 GeV with 6 (left) and 4 (right) APDs in the EM part of the calorimeter. 
The blue points and line in Figure~\ref{fig:had_en_response} show the resolution 
when the pion energy distribution is fitted by a Gaussian curve. The red ones, when the 
distribution is fitted by the Landau expression. We observe that the resolution is 
much better when the Landau fit is employed. 
It should be noted that the length of the tested calorimeter is only 4.26 interaction lengths
(almost a factor 3 smaller than the planned length of the final CASTOR calorimeter)
and thus there is considerable energy leakage at the end even at low pion energies. 
This does not permit an accurate estimation of the hadronic resolution.\\

\begin{figure*}[htbp] 
\begin{center}
\includegraphics[width=9cm,angle=-90]{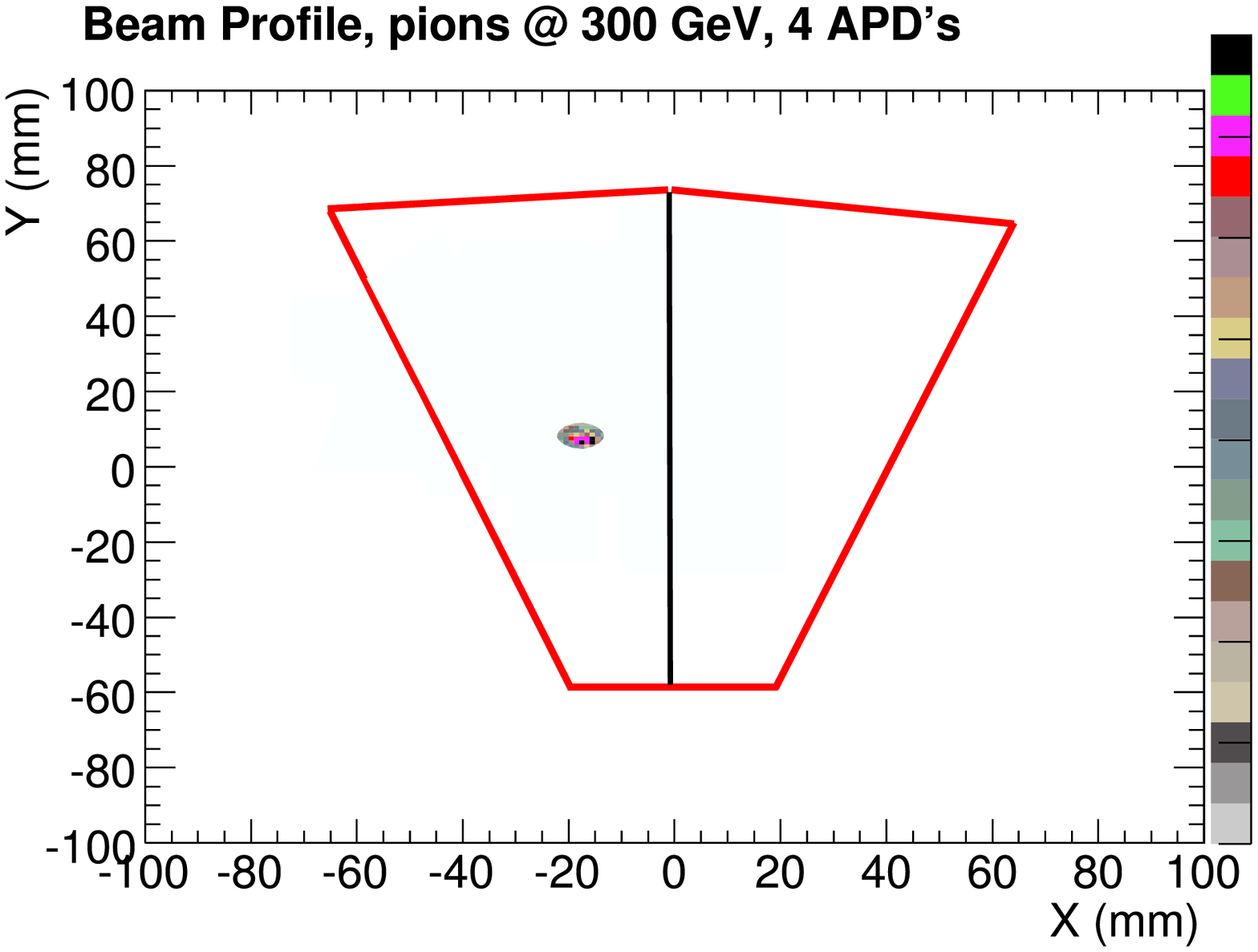}
\caption{Profile of the 300 GeV pion beam impinging on the left semi-octant region 
of the calorimeter.}
\label{fig:pion_hit}
\end{center}
\end{figure*}

\begin{figure*}[htbp] 
\begin{center}
\includegraphics[width=10cm,height=16cm,angle=-90]{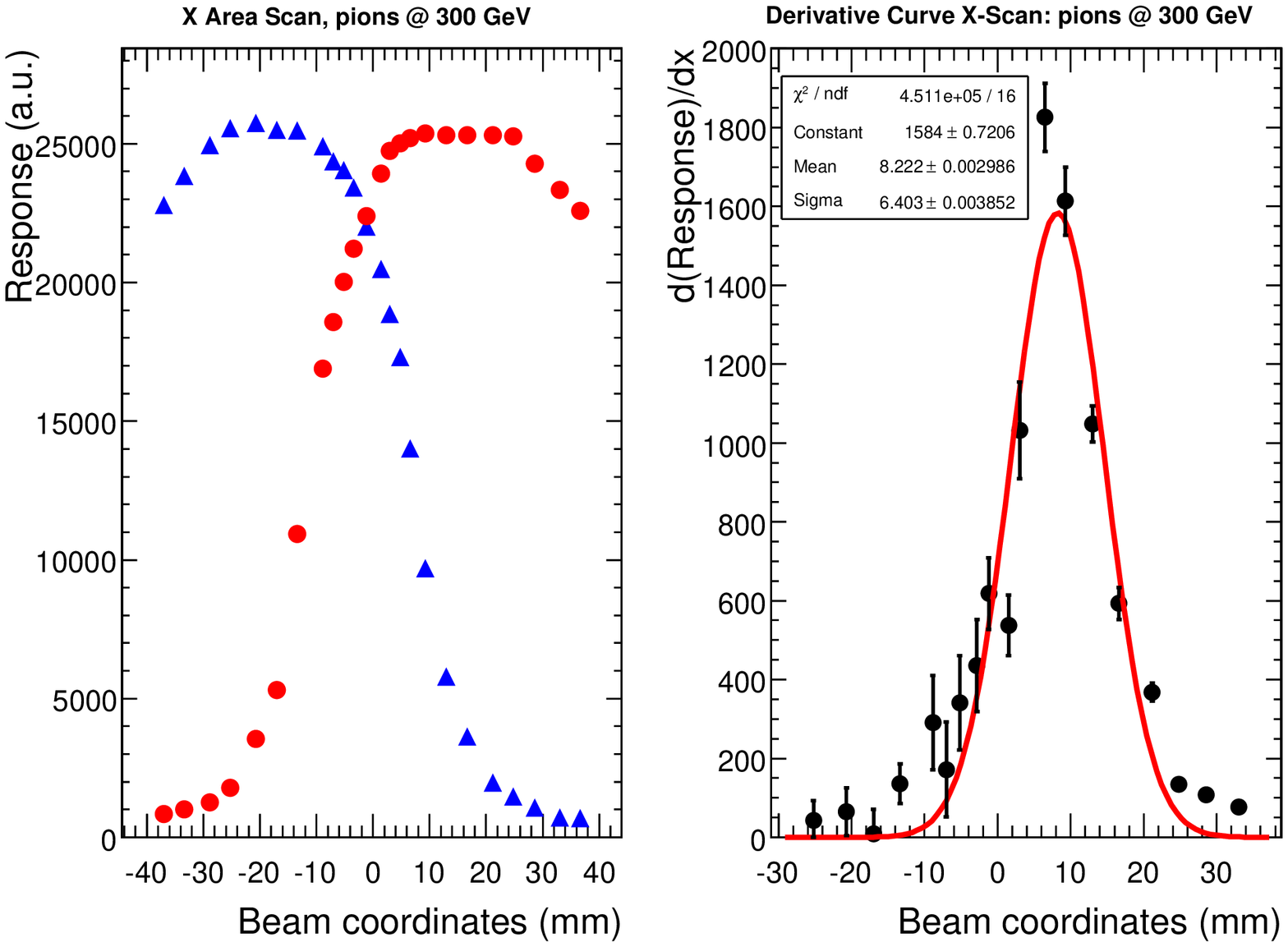}
\caption{$x-$scan along the face of the prototype for 300 GeV pions (left plot). 
The derivative of the sigmoid curve, giving the width of the hadronic 
shower distribution (right plot).}
\label{fig:had_x_scan}
\end{center}
\end{figure*}

\noindent {\bf Spatial Response: }
Figure~\ref{fig:pion_hit} shows the pion beam profile hitting the left 
semi-octant region of the prototype. We observe that the hadron beam is much more 
focused than the electron beam (see profile in Fig.~\ref{fig:elec_impact}). 
The spatial response of the prototype calorimeter to pions was obtained 
from the two EM semi-octant sectors, by moving the beam along the $x-$direction. 
The 1.03$\lambda_I$ inactive absorber was positioned in front of the calorimeter. 
The beam profile for each point was subdivided into a number of parts, 
each of diameter $\sim$5 mm, so that we obtained more impact points in total. 
Figure~\ref{fig:had_x_scan} shows the $x-$scan for pions of 300 GeV energy on the 
left and the derivative of this response with respect to $x$ on the right. 
The pion beam width has $\sigma_{HAD}$ = 6.4 mm, considerably larger than the 
corresponding electromagnetic one ($\sigma_{EM}$ = 1.7 mm, see 
Fig.~\ref{fig:em_esp_response}), as expected.

\section{Muon beam tests}
\label{sec:mu_beam}

\begin{figure*}[htbp] 
\begin{center}
\includegraphics[width=16.5cm]{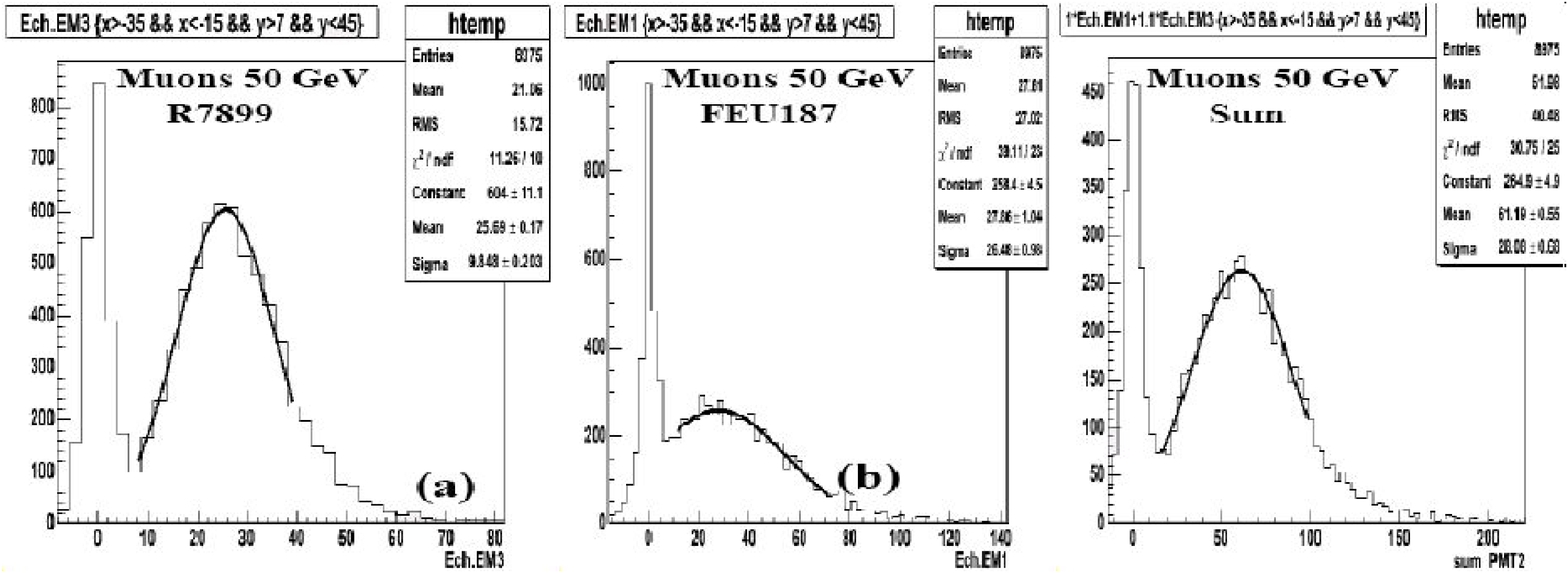}
\caption{Energy spectra measured in the EM section of prototype II with a
muon beam of 50 GeV energy and using two different PMTs: 
(a) Hamamatsu R7899, (b) RIE FEU187, and (c) the sum of both.}
\label{fig:muon_spec}
\end{center}
\end{figure*}

Muon energy spectra at 50 and 100 GeV were measured with the electromagnetic 
sector, using the PMT readout configuration. Figure~\ref{fig:muon_spec} shows the 
muon peak measured for the 50 GeV beam well separated from the pedestal at zero
counts. The lineshape has been obtained with two different PMTs: Hamamatsu R7899 
(Fig.~\ref{fig:muon_spec}a), and RIE FEU187 (\ref{fig:muon_spec}b). 
In Figure~\ref{fig:muon_spec}c, the sum of both EM readout units is shown.

From Figure~\ref{fig:muon_spec} we find that the Hamamatsu R7899 PMT performs 
much better than the RIE FEU187 one, in identifying the muon signal above the pedestal.
A disadvantage of  the R7899 PMT for this application is its large length, which prohibits its use,
even in the semi-octant geometry. 


\section{Monte Carlo simulation of prototype II}
\label{sec:MC}

\begin{figure*}[htbp] 
\begin{center}
\includegraphics[width=5cm]{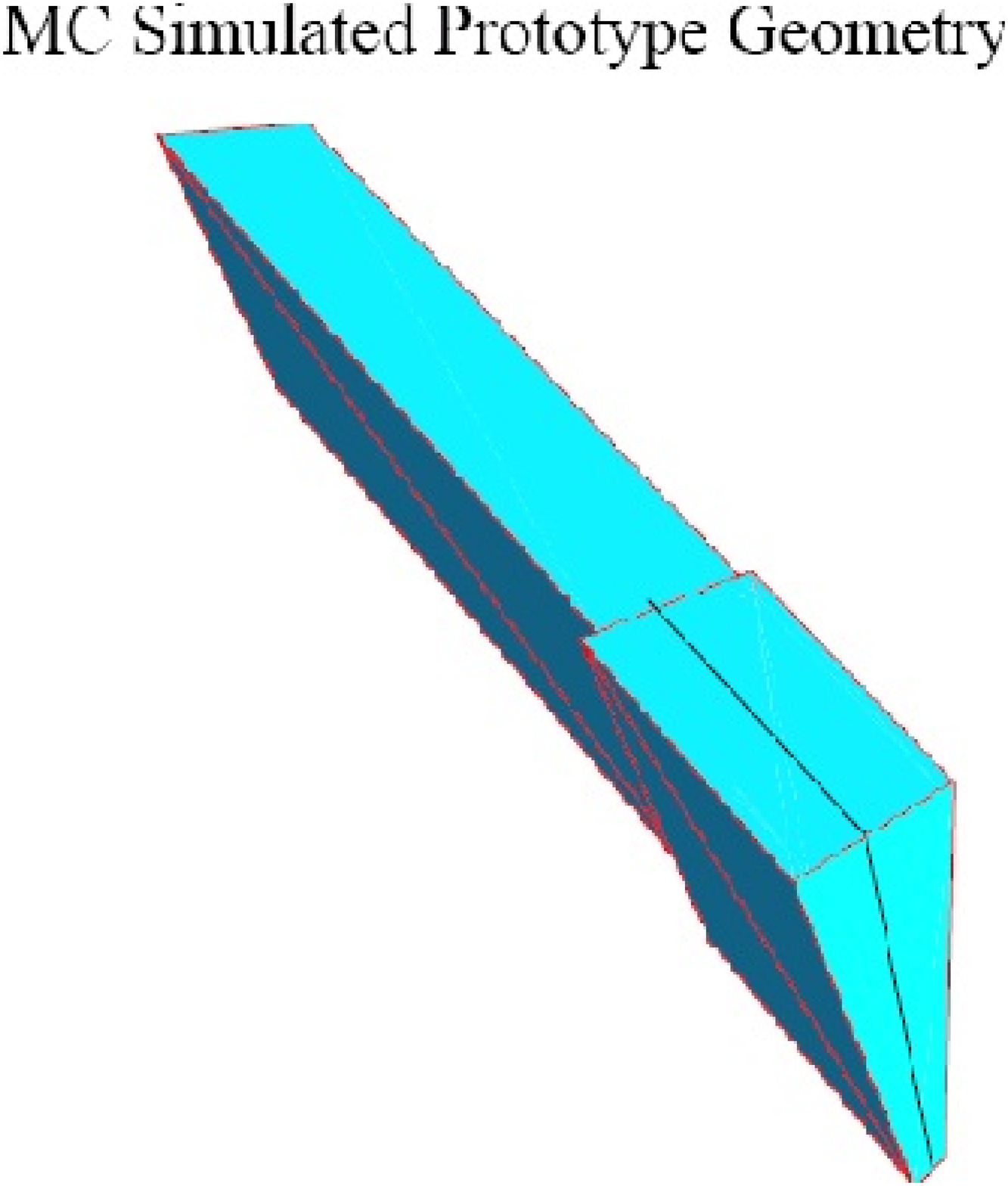}
\caption{Layout of the simulated geometry of the CASTOR prototype as implemented in GEANT4 (OSCAR 6.3.5).}
\label{fig:geant4_protoII}
\end{center}
\end{figure*}

Fig.~\ref{fig:geant4_protoII} shows the GEANT4~\cite{geant4} geometry of prototype II as 
implemented in the CMS software (OSCAR 6.3.5). The geometry of the electromagnetic section 
described in the simulations (\texttt{XML}-format) matches exactly that of the tested calorimeter.

\begin{figure*}[htbp] 
\begin{center}
\includegraphics[height=8.4cm,angle=-90]{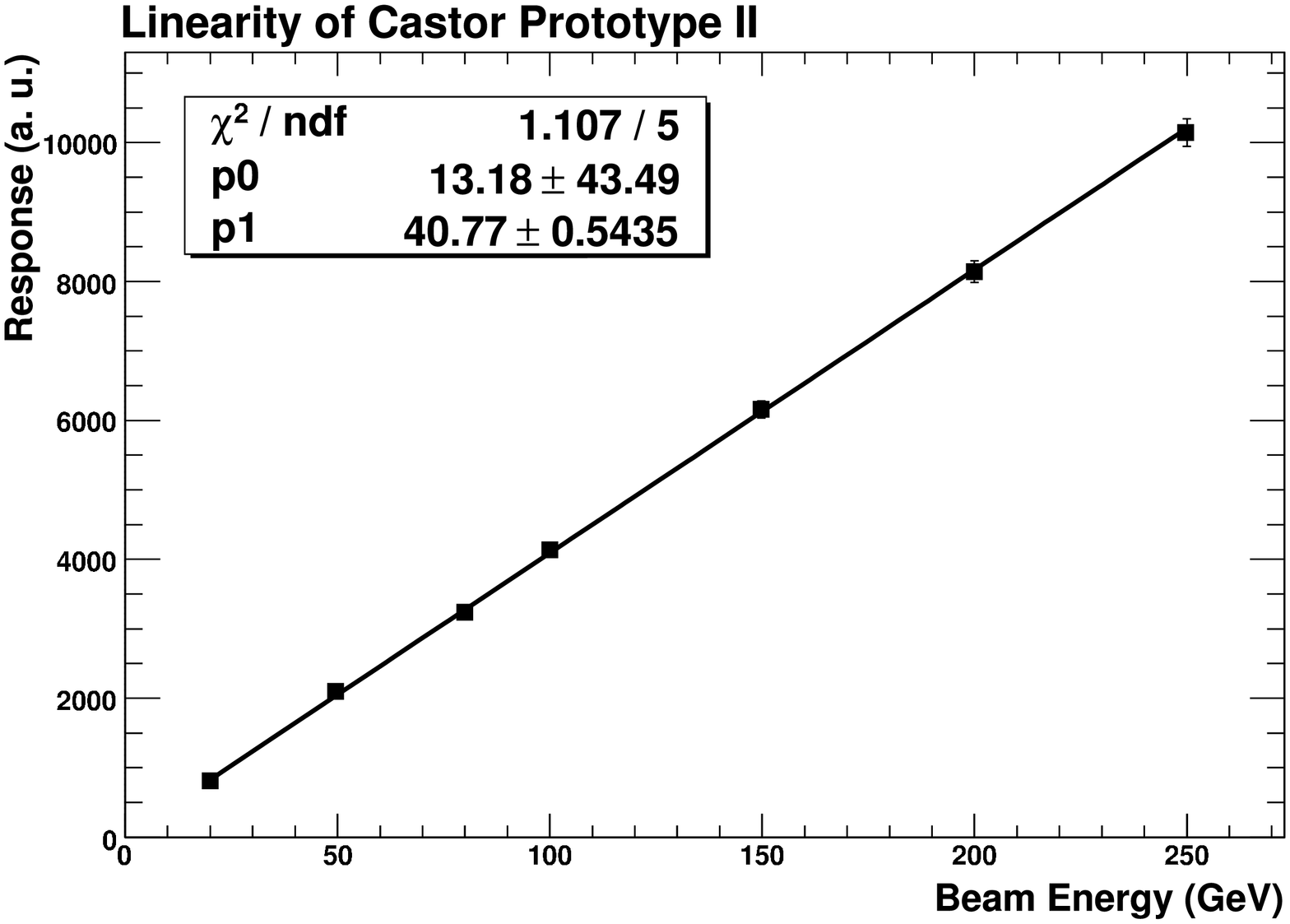}\hspace{-1cm}
\includegraphics[height=8.4cm,angle=-90]{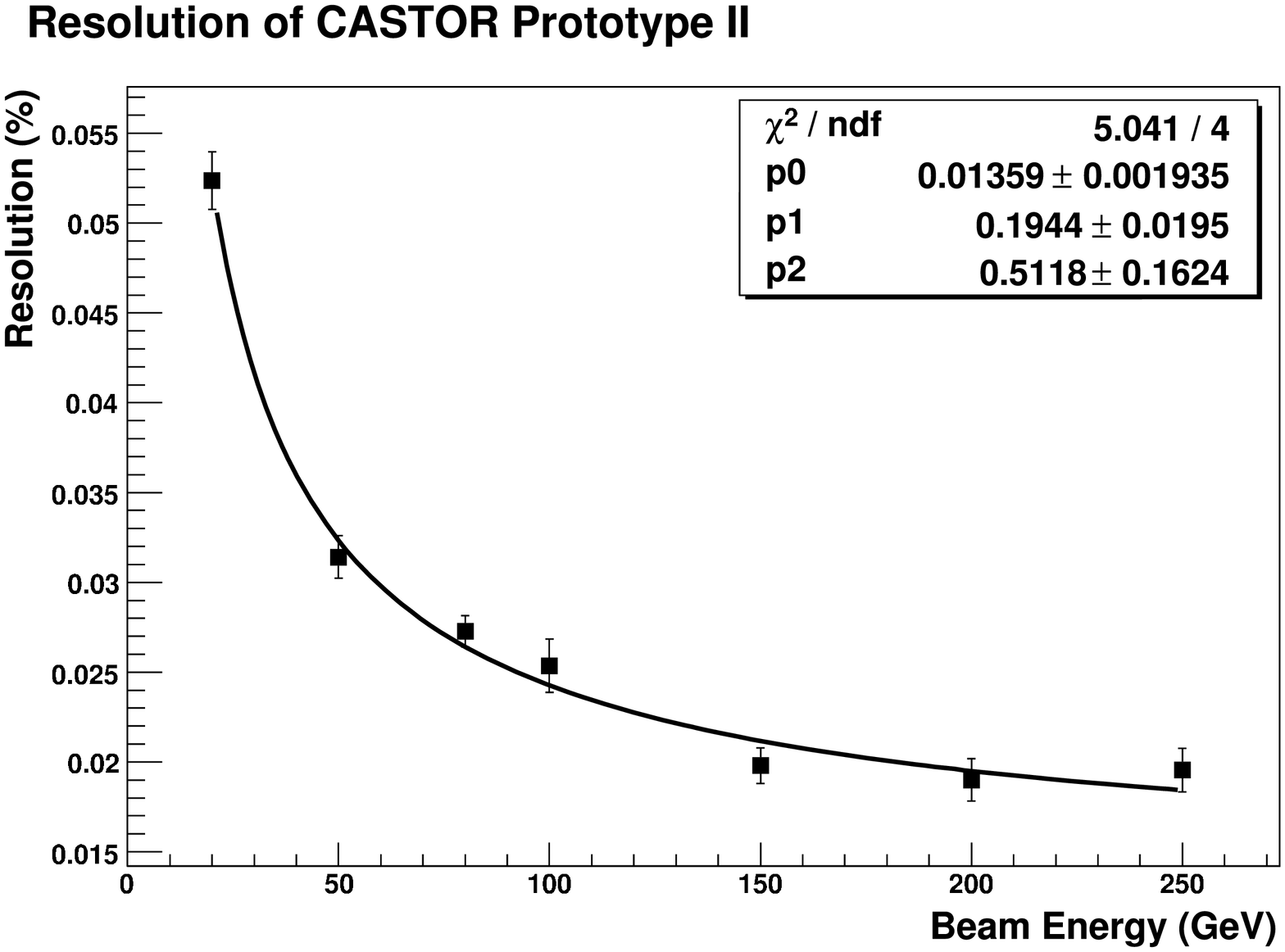}
\caption{Simulated energy response in terms of photo-electrons generated in the EM sections
of the CASTOR prototype: (a) linearity, (b) resolution.}
\label{fig:sim_en_response}
\end{center}
\end{figure*}

\begin{figure*}[htbp] 
\begin{center}
\includegraphics[height=12cm,angle=-90]{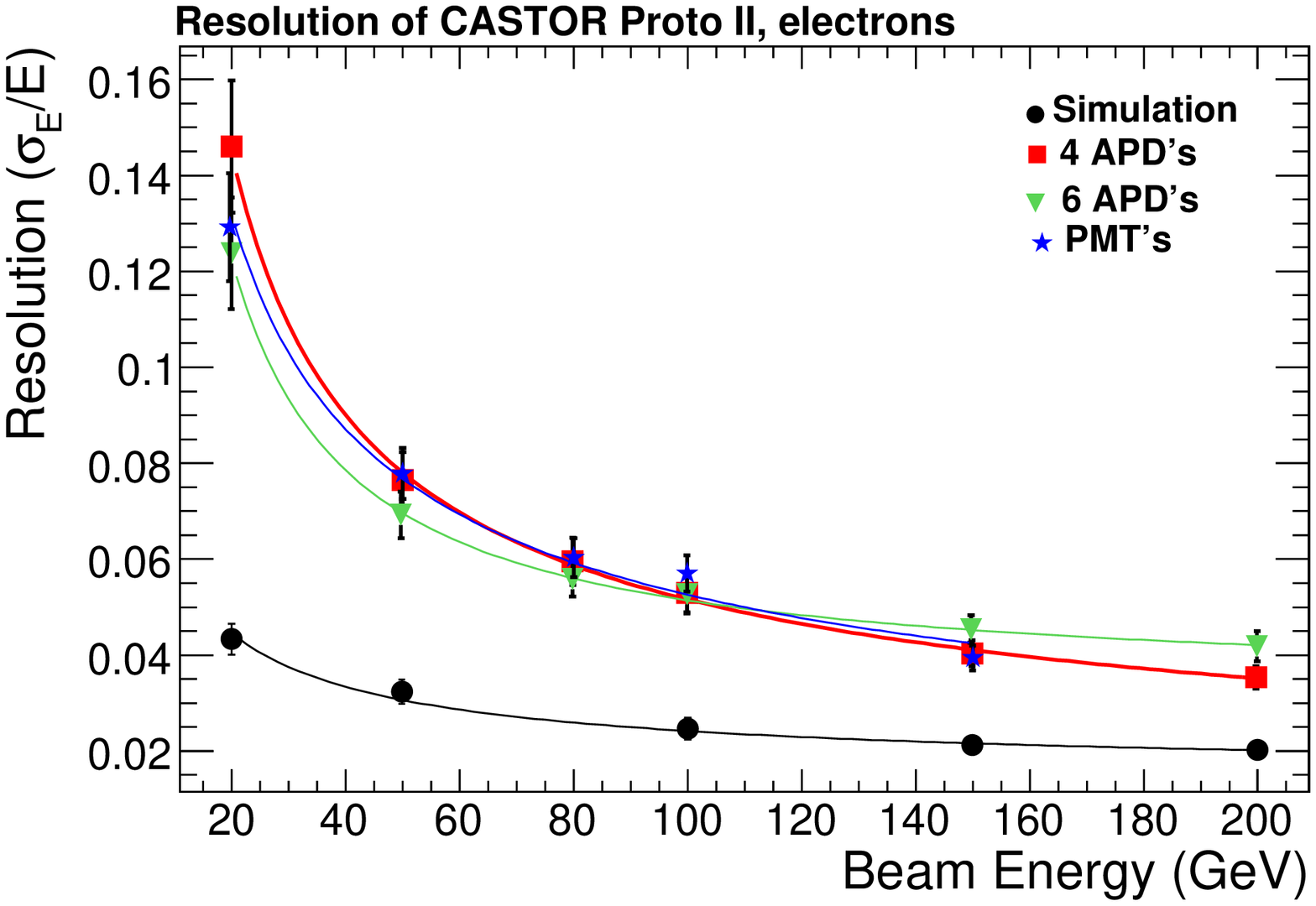}
\caption{Comparison of the experimental resolution for the three 
light readout configurations considered (4,6 APDs and PMTs) 
and the MC simulated one.}
\label{fig:comparison_data_sim}
\end{center}
\end{figure*}

We run simulations for 500 electron events with 7 different energies in the range $E$ = 20 -250 GeV
and studied the corresponding response in terms of the number of photoelectrons produced.
Figure~\ref{fig:sim_en_response} shows the simulated energy (a) linearity and (b) resolution 
of the prototype obtained assuming an overall efficiency (light transmission $\times$ quantum 
efficiency) of about 65\% for the APDs~\cite{castor_protoI}.
The linearity of the energy response is consistent with the experimental data (Fig.~\ref{fig:em_linearity}), but 
the energy resolution is 2--3 times better than the beam test results (Fig.~\ref{fig:comparison_data_sim}).

\begin{figure*}[htbp] 
\begin{center}
\includegraphics[height=15cm,angle=-90]{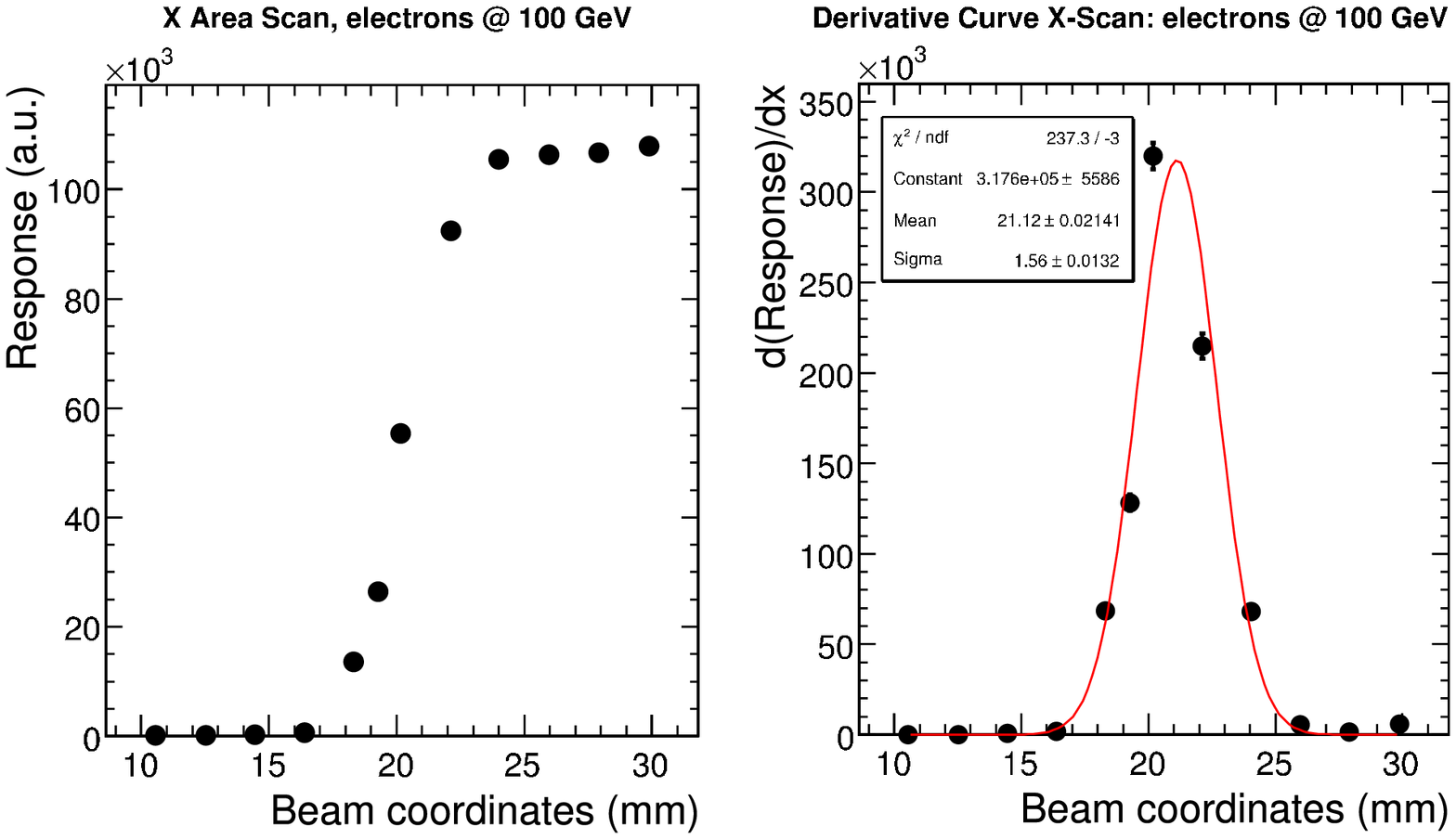}
\caption{(a) Simulated $x-$ profile of the electromagnetic shower. (b) Derivative of the simulated response
with respect to $x$, indicating the width of the EM shower.}
\label{fig:sim_xy_resol}
\end{center}
\end{figure*}

Figure~\ref{fig:sim_xy_resol} shows the $x$-spatial response of the electromagnetic 
shower simulated in GEANT4. In the MC simulation, the electron beam has a radius of 
1.5 mm, similar to the cut imposed in the analysis of the experimental data. The 
sigmoid curve is seen in Fig.~\ref{fig:sim_xy_resol}a and its $x$-derivative in 
Fig.~\ref{fig:sim_xy_resol}b, from which we obtain the width of 1.56 mm.
which is close to what one observes in the real data (Fig.~\ref{fig:em_esp_response}b)

\section{Summary}

We have presented a detailed performance study of the energetic and spatial responses of 
a second prototype of the CASTOR quartz-tungsten calorimeter of the CMS experiment. 
The results have been obtained from beam tests at CERN-SPS with high-energy electrons 
(20-200 GeV), pions (20-350 GeV) and muons (50, 150 GeV) and two different types of
photodetectors (APDs and PMTs) for the EM section of the calorimeter.
The main conclusions of this study can be summarized as follows:
\begin{enumerate}

\item EM Section: The semi-octant geometry has an efficient light-collection 
with 4 or 6 APDs. Due to the small height of the light-guide, 
a PMT readout can also be used, provided it is of small size. This has the advantage 
of higher gain (over the APD configurations), enabling the clear identification of the 
muon peak above the pedestal.

\item HAD Section: The octant geometry has an efficient light-collection for the hadronic 
section. However, the large height of the associated light-guides precludes this configuration 
in the limited space available for the CASTOR calorimeter in the very forward region of 
the CMS experiment.
\end{enumerate}

On the basis of physics concerns for both pp and heavy-ion interactions, the {\it semi-octant} 
geometry (which would correspond to 16 sectors covering full $\phi$) is, therefore, preferred. 
For this geometry, two reading-device options provide the desired performances: 
(i) 6 Hamamatsu-S8148 APDs per readout unit, and (ii) a small-size PMT, such as the 
RIE FEU-187. Both photodetectors should be tested/adapted for the radiation-harsh conditions 
of the CASTOR calorimeter ($\sim$10-100MGy accumulated, to be compared e.g. to the 
$\sim$3kGy expected for the CMS ECAL APDs~\cite{apd1}). The relative merits and difficulties 
of each option will be further studied in detail before a final decision is reached.


\section{Acknowledgments}

This work is supported in part by the Secretariat for Research of the University 
of Athens and by the Polish State Committee for Scientific Research (KBN) SPUB-M nr. 
620/E-77/SPB/CERN/P-03/DWM 51/2004-2006.  D.d'E. acknowledges support from 
the 6th EU Framework Programme (contract MEIF-CT-2005-025073).


\end{document}